\documentclass[prd,aps,floats,preprint,preprintnumbers,nofootinbib,showkeys]{revtex4-1}
\usepackage{graphicx}
\usepackage{latexsym}
\usepackage{amsmath,amssymb,mathrsfs,amscd}
\usepackage{feynmf}
\usepackage{graphics}
\usepackage{epsfig}
\usepackage{dsfont}
\usepackage{oldgerm}
\usepackage{color}
\usepackage{setspace}

\newcommand{\beq}{\begin{equation}}
\newcommand{\eeq}{\end{equation}}
\newcommand{\bea}{\begin{eqnarray}}
\newcommand{\eea}{\end{eqnarray}}

\newcommand{\A}{\mathcal{A}_{\theta}}

\newcommand{\Pa}{\mathbb{C}\mathscr{P}}

\newcommand{\Hpt}{H_\theta\mathscr{P}}
\newcommand{\M}{\mathcal{M}}
\newcommand{\B}{\mathscr{B}}
\newcommand{\pr}{\mathcal{P}_\alpha}
\newcommand{\dx}{{\rm d}}
\newcommand{\e}{{\rm e}}

\newcommand{\I}{\mathds{1}}
\newcommand{\R}{\mathbb{R}}

\newcommand{\Lr}{\mathscr{L}^\uparrow_+}
\newcommand{\Pg}{\mathscr{P}^\uparrow_+}
\newcommand{\Da}{\mathbb{C}\big(D^{(1)\infty}/D^{(1)\infty}_0\big)}

\newcommand{\PR}{\mathfrak{P}}

\newcommand{\Z}{\mathbb{Z}}
\newcommand{\F}{\mathscr{F}_\theta}

\begin{document}

\small
\preprint{SU-4252-913\vspace{1cm}} \setlength{\unitlength}{1mm}
\title{Quantum Geons and Noncommutative Spacetimes
\vspace{0.1cm}}
\author{ A. P.
Balachandran$^a$}\thanks{bal@phy.syr.edu}\author{A. Ibort$^b$}\thanks{albertoi@math.uc3m.es}\author{G. Marmo$^c$}\thanks{marmo@na.infn.it} \author{M. Martone$^{a,c}$}\thanks{mcmarton@syr.edu}
\affiliation{$^{a}$Department of Physics, Syracuse University, Syracuse, NY
13244-1130, USA\\
$^{b}$Departamento de Matem\'aticas, Universidad Carlos III de Madrid, 28911 Legan\'es, 
Madrid, Spain\\
$^{c}$Dipartimento di Scienze Fisiche, University of Napoli and INFN, Via Cinthia I-80126 Napoli, Italy}

\begin{abstract}
\begin{description}

\item[Dedication] {\it This article is dedicated to Josh Goldberg in appreciation of his friendship and his long years of service and contributions to the field of relativity.}
\begin{spacing}{1.8}
\end{spacing}
Physical considerations strongly indicate that spacetime at Planck scales is noncommutative. A popular model for such a spacetime is the Moyal plane. The Poincar\'e group algebra acts on it with a Drinfel'd-twisted coproduct, however the latter is not appropriate for more complicated spacetimes such as those containing Friedman-Sorkin (topological) geons. They have rich diffeomorphisms and mapping class groups, so that the statistics groups for {\it N} identical geons is strikingly different from the permutation group $S_N$.  We generalise the Drinfel'd twist to (essentially all) generic groups including finite and discrete ones, and use it to deform the commutative spacetime algebras of geons to noncommutative algebras. The latter support twisted actions of diffeomorphisms of geon spacetimes and their associated twisted statistics. The notion of covariant quantum fields for geons is formulated and their twisted versions are constructed from their untwisted counterparts.  {\it Non-associative} spacetime algebras arise naturally in our analysis. Physical consequences, such as the violation of Pauli's principle, seem to be one of the outcomes of such nonassociativity.

The richness of the statistics groups of identical geons comes from the nontrivial fundamental group of their spatial slices. As discussed long ago, extended objects like rings and {\it D}-branes also have similar rich fundamental groups. This work is recalled and its relevance to the present quantum geon context is pointed out.
 
\end{description}
\end{abstract}

\keywords{Quantum Field Theory, Noncommutative spacetime, Quantum Geons}

\maketitle

\section{Introduction}\label{sec:intro}

Attempts to localise Planck-scale spacetime volumes cause the formation of trapped surfaces and black holes by the principles of quantum theory and relativity. Therefore, as argued by Doplicher, Fredenhagen and Roberts (DFR) \cite{Doplicher}, general considerations suggest a limitation on the precision of spacetime measurements. 

Similar limitations in quantum theory such as the Heisenberg uncertainty relations $\Delta x\Delta p\gtrsim\hbar/2$ are accounted for by imposing commutation relations like $[x,p]=i\hbar$. We may thus speculate that limitations on spacetime measurements too can be incorporated by deforming the commutative algebra of functions on spacetime into a noncommutative algebra. For this algebra, for  spatial dimension $d=3$, rigorous four-volume ``uncertainty relations''  have  in fact been found by Doplicher {\em et al} \cite{Dop2}.

When spacetime is $\mathbb{R}^{d+1}$, a popular deformation is the Moyal algebra $\A(\mathbb{R}^{d+1})$ where the antisymmetric matrix $\theta=[\theta_{\mu\nu}]$ with constant real coefficients is the deformation parameter. Thus, the coordinate functions $\hat{x}_\mu$ in the algebra satisfy the commutation relations $[\hat{x}_\mu,\hat{x}_\nu]=i\theta_{\mu\nu}$.

Remarkably, despite the appearance of the constant matrix $\theta$, the group algebra $\Pa$ of the Hopf-Poincar\'e group $\mathscr{P}$ acts on $\A(\mathbb{R}^{d+1})$ as a Hopf algebra $H_\theta\mathscr{P}$ with coproduct \cite{chari,majid,aschieri}
\bea
&\Delta_\theta(g)=F^{-1}_\theta(g\otimes g)F_\theta, \quad F_\theta=\e^{-\frac{i}{2}P_\mu\otimes\theta^{\mu\nu}P_\nu}\in\Pa\otimes\Pa&\quad,\\
&P_\mu={\rm translation\ generators}&\quad.
\eea

The Moyal plane with its Hopf symmetry $\Hpt$ is a very particular deformation based on the Drinfel'd twist $F_\theta\in\Pa\otimes\Pa$. It has the important physical property that its deformed quantum field theories (qft's) can be obtained from the undeformed qft's based on the commutative algebra $\mathcal{A}_0(\R^{d+1})$, the map from one to the other involving a known ``dressing transformation'' \cite{dress1,dress2}. Generic deformations do not support symmetries of $\mathcal{A}_0(\R^{d+1})$ in any sense. Even when they do, they need not admit unitary qft's. A simple example is the Wick-Voros plane \cite{Mario,Mario3}.

The spatial slice $\R^d$ is not the only admissible spatial slice for asymptotically flat spacetimes. Friedman and Sorkin \cite{Sorkin} have studied generic asymptotically flat spatial slices and have come up with their remarkable interpretation in terms of gravitational topological excitations called ``topological'' or ``Friedman-Sorkin'' ``geons''. The diffeomorphisms (diffeos) of geon spacetimes are much richer than those from the topologically trivial ones. In particular, they contain discrete subgroups encoding the basic physics of geons. It was a striking discovery of Friedman and Sorkin that the geon spin even in pure gravity can be $1/2$ or its odd multiples \cite{Sorkin,Fried,Geonbal,Witt}. The statistics groups of identical geons are also novel. Their precise identification requires further considerations as we shall see.

In this paper, we develop a machinery to construct Drinfel'd twists for generic and in particular discrete diffeos. We then recall the notion of covariant quantum fields well-studied for $\mathcal{A}_0(\R^{d+1})$ and extend them for generic spacetimes \cite{covariant}. This helps us construct covariant twisted fields for geons using the above twists. The requirement of covariance puts conditions on acceptable twists for quantum fields and eliminates many. Previously it eliminated the Wick-Voros twist \cite{Mario,Mario3}.

Spacetimes emergent from these twists are noncommutative as is appropriate at geon scales according to DFR \cite{Doplicher}. There is a diffeo-invariant way to define the size of a geon \cite{Kauff} and it is expected to be of Planck-scale. Spacetime noncommutativity emergent from our approach is localised at geons and is of this scale just as we wish for. 

As we indicate, several novel spacetimes including non-associative spacetimes and new sorts of statistics algebras are indicated by this work. In this paper, these matters are discussed only in a preliminary manner. But already, new phenomena like non-Pauli transitions are suggested as we will see. 

It was recognised long ago \cite{Kauff,Sum} that the novel statistics groups of geons are a reflection of their being extended excitations like solitons, and that similar statistics groups occur for other extended objects like rings and {\it D}-branes. Our geon considerations can be adapted to their cases too. We recall these older results briefly towards the end of the paper. Maybe, they will allow explicit computations of physical effects from Drinfel'd twists, and give us indications on what to expect for geons. Detailed calculations with geons are hard as they involve quantum gravity (see however \cite{Sorkin,Fried}).

\section{What are geons}

\begin{figure}
\includegraphics[scale=.38]{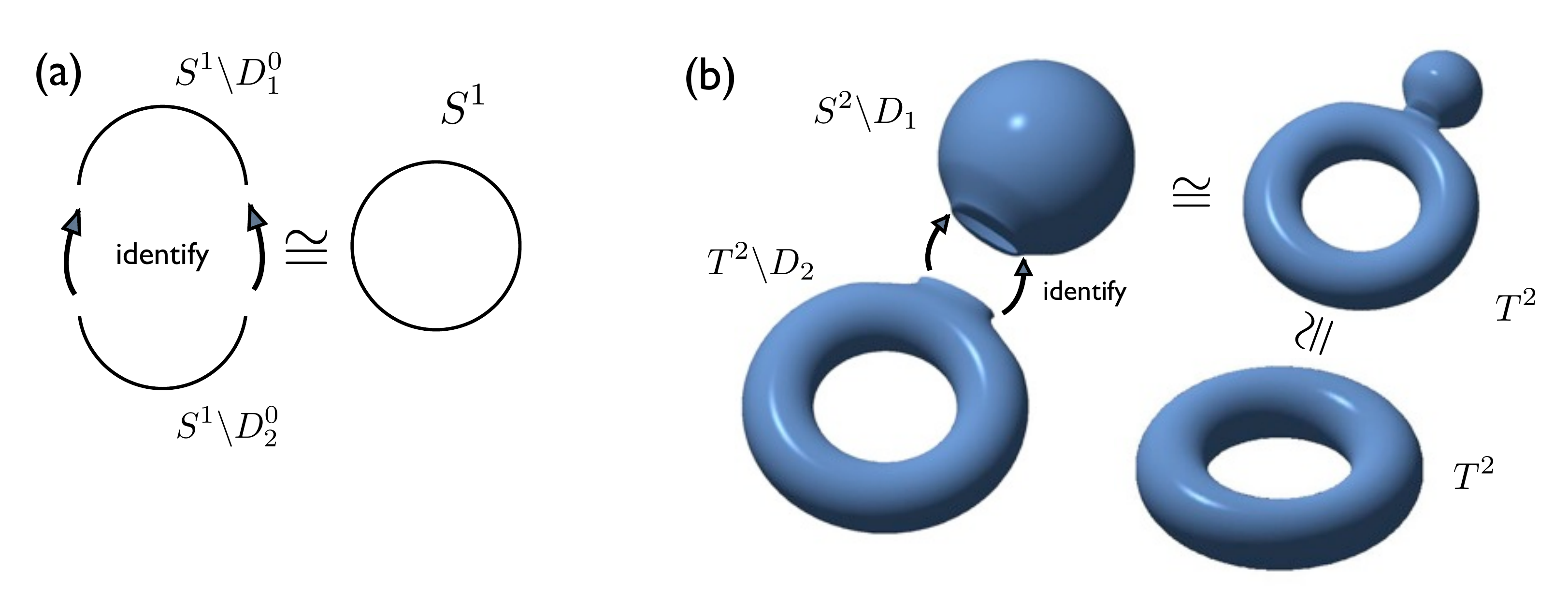} 
\caption{(a) $S^1\#S^1 \cong S^1$. (b) $T^2\#S^2 \cong T^2 \cong S^2\#T^2$}
\label{fig1}
\end{figure}

This is a short review section on the topology of low--dimensional manifolds leading up to those which support geons (``geon manifolds''). The original literature is best consulted for detailed information \cite{Sorkin,Fried,Geonbal,Witt}.

Given two closed (compact and boundary--less) manifolds $\mathcal{M}_1$ and $\mathcal{M}_2$ of dimension $d$, their connected sum $\mathcal{M}_1\#\mathcal{M}_2$ is defined as follows. Remove two balls $\mathscr{B}_1$ and $\mathscr{B}_2$ from $\mathcal{M}_1$ and $\mathcal{M}_2$, leaving two manifolds $\mathcal{M}_i\backslash\mathscr{B}_i$ with spheres $S^{d-1}_i$ ($S^{d-1}_i\sim S^{d-1}$) as boundaries $\partial(\mathcal{M}_i\backslash\mathscr{B}_i)$. Then $\mathcal{M}_1\#\mathcal{M}_2$ is obtained by identifying these spheres. If $\mathcal{M}_i$ are oriented, this identification must be done with orientation-reversal so that $\mathcal{M}_1\#\mathcal{M}_2$ is oriented. 

Connected summing, $\#$,  is associative and commutative:
\begin{itemize}
\item[a)] $\mathcal{M}_1\#(\mathcal{M}_2\#\mathcal{M}_3) \cong (\M_1\#\M_2)\#\M_3$ so that we can write $\M_1\#\M_2\#\M_3$;

\item[b)]  $\M_1\#\M_2 \cong \M_2\#\M_1$.
\end{itemize}

\begin{figure}
\includegraphics[scale=.3]{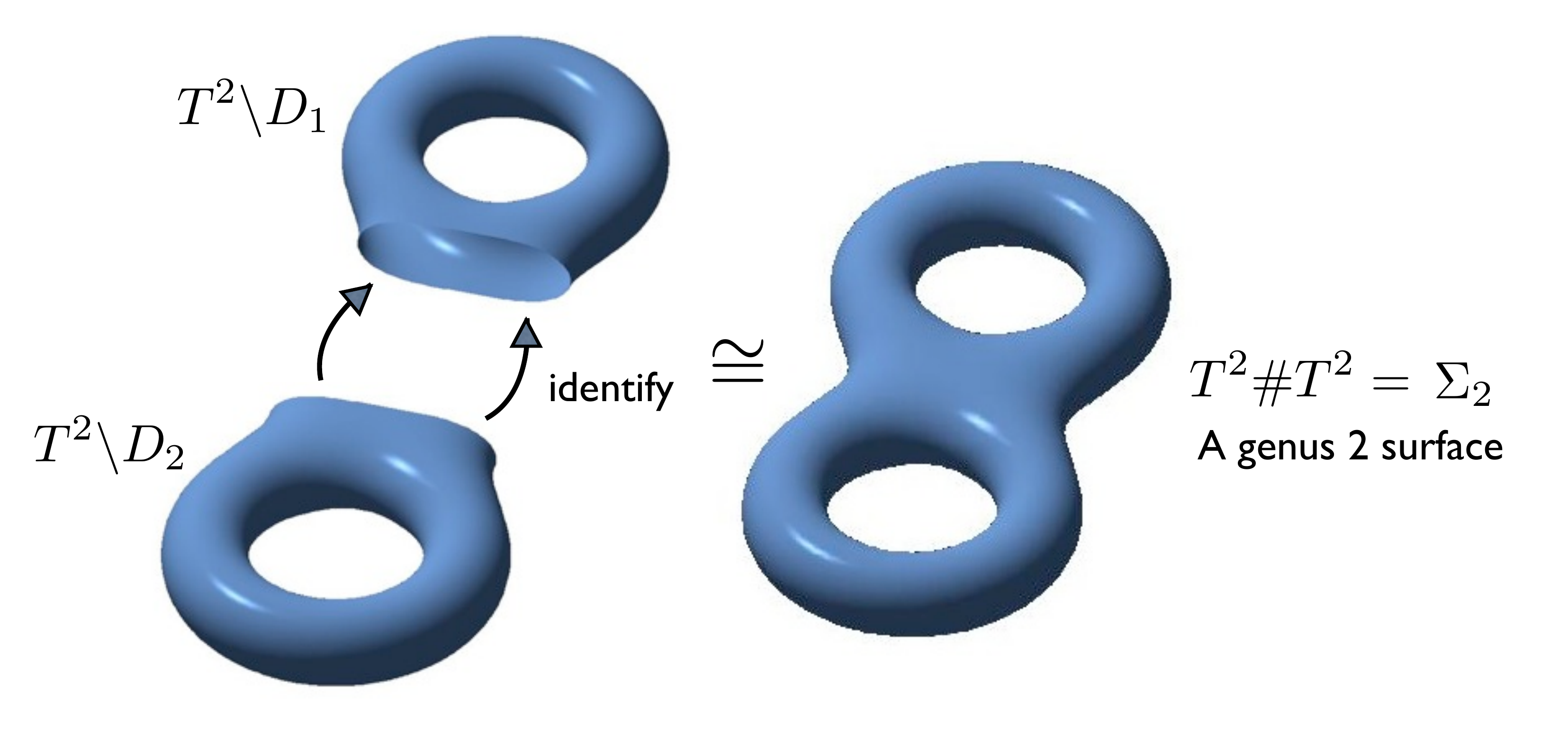} 
\caption{$T^2\#T^2 \cong \Sigma_2$. Genus 2 surface.}
\label{fig3}
\end{figure}

Here are some simple examples:

\begin{itemize}

\item $d=1$.  $S^1\#S^1 \cong S^1$.  (See Fig. \ref{fig1} (a)).

\item $d=2$.  $S^2\#S^2 \cong S^2$. 

\item $d=2$.  $T^2\#S^2=T^2 \cong S^2\#T^2$. (See Fig. \ref{fig1} (b)).

\item $d=2$.   $T^2\#T^2 \cong \Sigma_2$. Genus two manifold. (See Fig. \ref{fig3}).

\end{itemize}


As the examples here suggest, for any dimension $d$, $\M\#S^d \cong S^d\#\M \cong \M$.

These considerations can be extended to asymptotically flat manifolds. If $\M_1$ is asymptotically flat and $\M_2$ is closed and both are oriented (and of the same dimension), then $\M_1\#\M_2$ is obtained by removing balls $\B_i$ from $\M_i$ and identifying the boundaries $\partial(\M_i\backslash\B_i)$ compatibly with orientation as pointed out above.  The connected sum $\M_1\#\M_2$ is asymptotically flat and oriented. 

We will now state certain basic results in low--dimensional topology considering only closed or asymptotically flat, and oriented manifolds $\M$. In the asymptotically flat case, we will insist that there is only one asymptotic region. That is, the asymptotic region of $\M$ is homeomorphic to the complement of a ball $\B^d$ in $\R^d$. In other words $\M$ has one asymptotic region if all its topological complexities can be encompassed within a sphere $S^{d-1}\subset\M$.

The case $d=1$ is trivial, there being only two such manifolds $S^1$ and $\R^1$. ($\R^1$ has ``one'' asymptotic region in the above sense even though it is not connected.)

The basic results of interest for $d=2$ and 3 are as follows.

\subsection{{\it Closed Manifolds}}

In $d=2$ and 3, there is a class of special closed manifolds called {\it prime} manifolds. Any closed manifold $\M\neq S^d$ for $d=2$ or 3 is a {\it unique} connected sum of prime manifolds $\pr$ (with the understanding that spheres are not inserted in the connected sum):
\beq\label{geo1}
\M=\#_\alpha\pr\, .
\eeq
(All manifolds have the same dimension. If $\M=S^d$, then (\ref{geo1}) is substituted by the triviality $S^d=S^d$,
hence, a better way to write (\ref{geo1}) is $\M=\#_\alpha\pr\ {\rm mod}\ S^d$.) 

The uniqueness of (\ref{geo1}) implies that a prime $\pr$ cannot be decomposed as the connected sum of two or more primes. (It is indecomposable just like a stable elementary particle.)

For $d=2$, there is just one prime, namely the torus. In that case, $T^2\#T^2\#...\#T^2$ with $k$ terms is just a genus $k$ surface (see Fig. \ref{fig3} for $k = 2$). 

For $d=3$, there are an infinity of prime manifolds. They are not fully known. Representative examples are the following:
\begin{itemize}
\item[a)] {\it Spherical Space Forms}. Notice that $S^3 \cong SU(2)$ by writing 
\beq
SU(2)\ni g=
\left(
\begin{array}{cc}
\xi_1&-\bar{\xi}_2\\
\xi_2&\bar{\xi}_1
\end{array}
\right),\ |\xi_1|^2+ |\xi_2|^2=1\, .
\eeq
Then $SO(4)=\frac{SU(2)\times SU(2)}{\mathbb{Z}_2}$ acts on $S^3$ by 
\beq
g\to hgh'^{-1},\quad h,h'\in SU(2)\, .
\eeq
There are several discrete subgroups of $SO(4)$ which act freely on $S^3$. Such free actions are given for example by the choices $h\in\mathbb{Z}_p$, $h'\in\mathbb{Z}_q$ where $p$ and $q$ are relatively prime. The quotients of $S^3$ by the free actions of discrete subgroups of $SO(4)$ are called spherical space forms. For the above example with cyclic groups $\mathbb{Z}_{p,q}:=\mathbb{Z}_p\times\mathbb{Z}_q$ the quotients are Lens spaces $L_{p,q}$ \cite{Homo}. Of these $L_{1,2}$ and $L_{2,1}$ are $\mathbb{R} P^3$.

Spherical space forms are prime and admit metrics with constant positive curvature. They have been studied exhaustively from the point of view of quantum gravity by Witt \cite{Witt}.

\item[b)] {\it Hyperbolic spaces}. Consider the hyperboloid 
\beq
\mathcal{H}^+:\{x= (x_0, \vec{x} )\in\R\times \R^3 \cong \R^4:(x_0)^2-(\vec{x})^2=1, x_0 > 0\}
\eeq
in $\mathbb{R}^4$. The connected Lorentz group $\Lr$ acts transitively on $\mathcal{H}^+$. Let $D\subset\Lr$ be a discrete subgroup acting freely on $\mathcal{H}^+$.  Then $\mathcal{H}^+/D$ is a hyperbolic space.

Hyperbolic spaces are prime and admit metrics with constant negative curvature. 
\end{itemize}

There are other primes as well such as $S^2\times S^1$ which do not fall into either of these classes.

\subsection{{\it Manifolds with one asymptotic region}}

These manifolds $\M_\infty$ also have a unique decomposition of the form
\beq
\M_\infty=\R^d\#_\alpha\pr
\eeq
where $\pr$ are the prime manifolds we discussed previously.
Manifolds with one asymptotic region can be obtained from closed manifolds $\M$ by removing a point (``point at $\infty$'').

\section{On Diffeos}

Spatial manifolds of interest for geons are $\M_\infty$. They serve as Cauchy surfaces in globally hyperbolic spacetimes. Spacetime topology is taken to be $\M_\infty\times\R$ where $\R$ accounts for time.

In the standard Drinfel'd twist approach, the twist $F_\theta$ belongs to $\mathbb{C}\mathscr{G}\otimes\mathbb{C}\mathscr{G}$, where $\mathscr{G}$ represents the symmetry group which in relativistic quantum field theory is taken to be $\mathscr{P}$ or its identity component $\Pg$. In order to let the twist act on a geon spacetime we should identify the substitute for $\mathscr{P}$ or $\Pg$. To achieve that we need to recall a few properties of quantization of diffeomorphism-invariant theories. We will present a summary of the main ideas here. For a self-contained treatment of the topic we refer the reader to \cite{book}.

It is a result of quantization on multiply connected configuration spaces \cite{Spara} that there is an action on the Hilbert space $\mathcal{H}$ of $\pi_1(\mathcal{Q})$, where $\mathcal{Q}$ is the configuration space of the classical system we want to quantize. This action can also be shown to commute with the action of any observable on $\mathcal{H}$. Now $\mathcal{H}$ can be decomposed into the direct sum $\mathcal{H}\cong\bigoplus\mathcal{H}^{(l)}$ of carrier spaces of irreducible representation of $\pi_1(\mathcal{Q})$.   (More precisely this is so only if $\pi_1(\mathcal{Q})$ is abelian. Otherwise $\mathcal{H}$ carries only the action of the center of the group algebra $\mathbb{C}\pi_1(\mathcal{Q})$, see for instance \cite{book}.)  Since all the observables commute with the action of $\pi_1(\mathcal{Q})$, they take each $\mathcal{H}^{(l)}$ into itself. These quantizations for different $l$ are generally inequivalent. In other words each $\mathcal{H}^{(l)}$ provides an inequivalent quantization of the classical system \cite{book}. These results have been widely used from molecular physics to quantum field theory. The $\theta$ angle of QCD it is in fact understood in such a topological way. 

In a theory of quantum gravity we consider $\pi_1(\mathcal{Q})$ as the group to twist. We turn now into the study of what this group looks like.

In general relativity the configuration space is very different from the usual $\mathbb{R}^{3n}$, as it is in the $n$-particle case. Specifically it is constructed from the set of all possible Riemannian metrics on a given space-like Riemannian manifold $\M$, which we will indicate as Riem$(\M)$. We also require, in order to make sense of concepts constantly used in physics like energy, that $\M$ is asymptotically flat. So we restrict $\M$ to what has been called $\M_\infty$ above.  We indicate by Riem$(\M_\infty)$ the space of metrics on it. 

Not all possible metrics on $\M_\infty$ represent physically inequivalent ``degrees of freedom'' though. Because of diffeomorphism invariance we should consider only Riem$(\M_\infty)$ upto the action of $D^\infty$, the diffeos which act trivially at infinity. We thus find for the configuration space $\mathcal{Q}$ of general relativity: $\mathcal{Q}\equiv{\rm Riem}(\M_\infty)/D^\infty$.

The next step is to compute the fundamental group of $\mathcal{Q}$. We first quote the result:
\beq\label{geo32}
\pi_1\Big({\rm Riem}(\M_\infty)/D^\infty\Big)=D^\infty/D^\infty_0:={\rm MCG}(\M_\infty)
\eeq
where $D^\infty_0$ is the (normal) subgroup of $D^\infty$ which is connected to the identity and MCG denotes the Mapping Class Group.  This  group is an important invariant of topological spaces.

Here is the proof of (\ref{geo32}). It can be shown that the action of $D^\infty$ on Riem$(\M_\infty)$ is free. Thus $\mathcal{Q}$ is the base manifold of a principal bundle Riem$(\M_\infty)$ with structure group $D^\infty$. By a well-known theorem of homotopy theory \cite{Steenrod}, the following sequence of homotopy groups is then exact:
\beq\label{geo31}
...\to\pi_1\Big({\rm Riem}(\M_\infty)\Big)\to\pi_1(\mathcal{Q})\to\pi_0(D^\infty)\to\pi_0\Big({\rm Riem}(\M_\infty)\Big)\to...
\eeq
As the space of Riemmanian asymptotically flat metrics is topologically ``trivial'', that is $\pi_n\Big({\rm Riem}(\M_\infty)\Big)\equiv\I,\ \forall n$, (\ref{geo31}) becomes:
\beq
\I\to\pi_1(\mathcal{Q})\to\pi_0(D^\infty)\to\I
\eeq
from which (\ref{geo32}) follows.

The nontrivial structure of the MCG$(\M_\infty)$ leads to striking results like the possibility of spinorial states from pure gravity. 
Let us briefly discuss this interesting result.  
The group $D^\infty$ contains a diffeo called the $2\pi$-rotation diffeo $R_{2\pi}$. It becomes a $2\pi$ rotation on quantum states. It may or may not be an element of $D^\infty_0$.   Now ``the momentum constraints'' of general relativity imply that $D^{\infty}_0$ acts as identity on all quantum states. Thus it is only the group $D^\infty/D^\infty_0$ (or more generally $D/D^\infty_0$ where $D$ may contain elements which asymptotically act  for example as rotations and translations) which can act nontrivially on quantum states. The conclusion in the following relies on this fact.

If $R_{2\pi}\in D^\infty_0$ then it maps to the identity in $D^\infty/D^\infty_0$ and on quantum states.

If $R_{2\pi}\notin D^\infty_0$, then it does not map to identity in $D^\infty/D^\infty_0$ and can act nontrivially on quantum states.

For $d\geq3$, $R^2_{2\pi}$ is always in $D^\infty_0$ and hence always acts trivially on quantum states.

Thus if $d\geq3$ and $R_{2\pi}\notin D^\infty_0$, there can exist quantum geons with $2\pi$ rotation=$-\I$ on their Hilbert space. In fact suppose that $\psi\in\mathcal{H}$ is a physical state on which $R_{2\pi}$ does not act trivially, $\hat{R}_{2\pi}\psi\neq\psi$. But $R_{4\pi}=R^2_{2\pi}$ acts trivially on $\mathcal{H}$. Then the state $\psi':=(\frac{\I-\hat{R}_{2\pi}}{2})\psi$ is spinorial:
\beq
\hat{R}_{2\pi}\psi'=\frac{1}{2}(\hat{R}_{2\pi}-\hat{R}_{4\pi})\psi=-\psi'\quad.
\eeq

It was a remarkable observation of Friedman and Sorkin \cite{Sorkin,Fried,Geonbal} that there exist primes $\pr$ such that $R_{2\pi}\notin D^\infty_0$ for $\M=\R^d\#\pr$. These are the ``spinorial'' primes. The quantisation of the metric of such $\R^3\#\pr$ can lead to vector states with spin $\frac{1}{2}+n$ ($n\in\mathbb{Z}^+$). Thus we can have ``spin $\frac{1}{2}$ from gravity''.

For $d=2$, the situation is similar, but $R^2_{2\pi}$ or any nontrivial power of $R_{2\pi}$, need not be in $D^\infty_0$. That is indeed the case for $\R^2\#T^2$ \cite{Anez}. That means that the quantum states for such geon manifolds can have fractional spin, can be anyons. 


\subsection*{{\it Notation.}}

Here we introduce some notation. We will call the diffeo groups of $\M_\infty=\R^d\#\pr$ which are asymptotically Poincar\'e, asymptotically identity and the component connected to the identity of the latter as $D^{(1)}$, $D^{(1)\infty}$ and $D^{(1)\infty}_0$ respectively. We will also refer to $D^{(1)\infty}/D^{(1)\infty}_0$ as the internal diffeos of the prime $\pr$. Similarly $D^{(N)}$,$D^{(N)\infty}$ and $D^{(N)\infty}_0$ will refer to the corresponding groups in the case of $N$-geon manifolds $\M_\infty=\R^d\#\pr\#...\#\pr$, where the primes are all the same. They are appropriate for constructing vector states of several identical geons.

The MCG of an $N$-geon manifold can be decomposed into semi-direct products involving three groups:
\beq\label{geo36}
D^{(N)\infty}/D^{(N)\infty}_0\equiv\Big(\mathscr{S}\rtimes\left[\times^ND^{(1)\infty}/D^{(1)\infty}_0\right]\Big)\rtimes S_N\quad.
\eeq
Here $A\rtimes B$ indicates the semi-direct product of $A$ with $B$ where $A$ is the normal subgroup.

In the above we could remove the brackets as it has been shown in \cite{MCG1,MCG2} that the above semi-direct product is associative.

The last two factors in (\ref{geo36}) are easily understood, the second term being the $N$-th direct product of the MCG of the single geon manifold $\M_\infty=\R^d\#\pr$ and $S_N$ being the usual permutation group of $N$ elements  that consists of elements which permute the geons. 

The first term, namely $\mathscr{S}$, is called the group of ``slides'' and consists of diffeos which take one prime through another along non-contractible loops. The existence of such a term is strictly linked with the fact that the primes are not simply connected. In fact elements of $\mathscr{S}$ can be described using elements of fundamental groups of the single primes $\pr$. Since we are not interested in the full details of the MCG, we refer the reader to the literature for further details \cite{MCG1,MCG2} while we now move on to the analysis of the $N=2$ case where we can also get a better understanding of what slides represent.

As we said the group $D^{(2)\infty}/D^{(2)\infty}_0$ of the manifold $\R^d\#\pr\#\pr$ appropriate for two identical geons contains diffeos corresponding to the exchange $E^{(2)}$ of geons and a new type of diffeos called slides besides the diffeos $D^{(1)\infty}/D^{(1)\infty}_0$ of $\R^d\#\pr$.

\begin{figure} 
\includegraphics[scale=0.4]{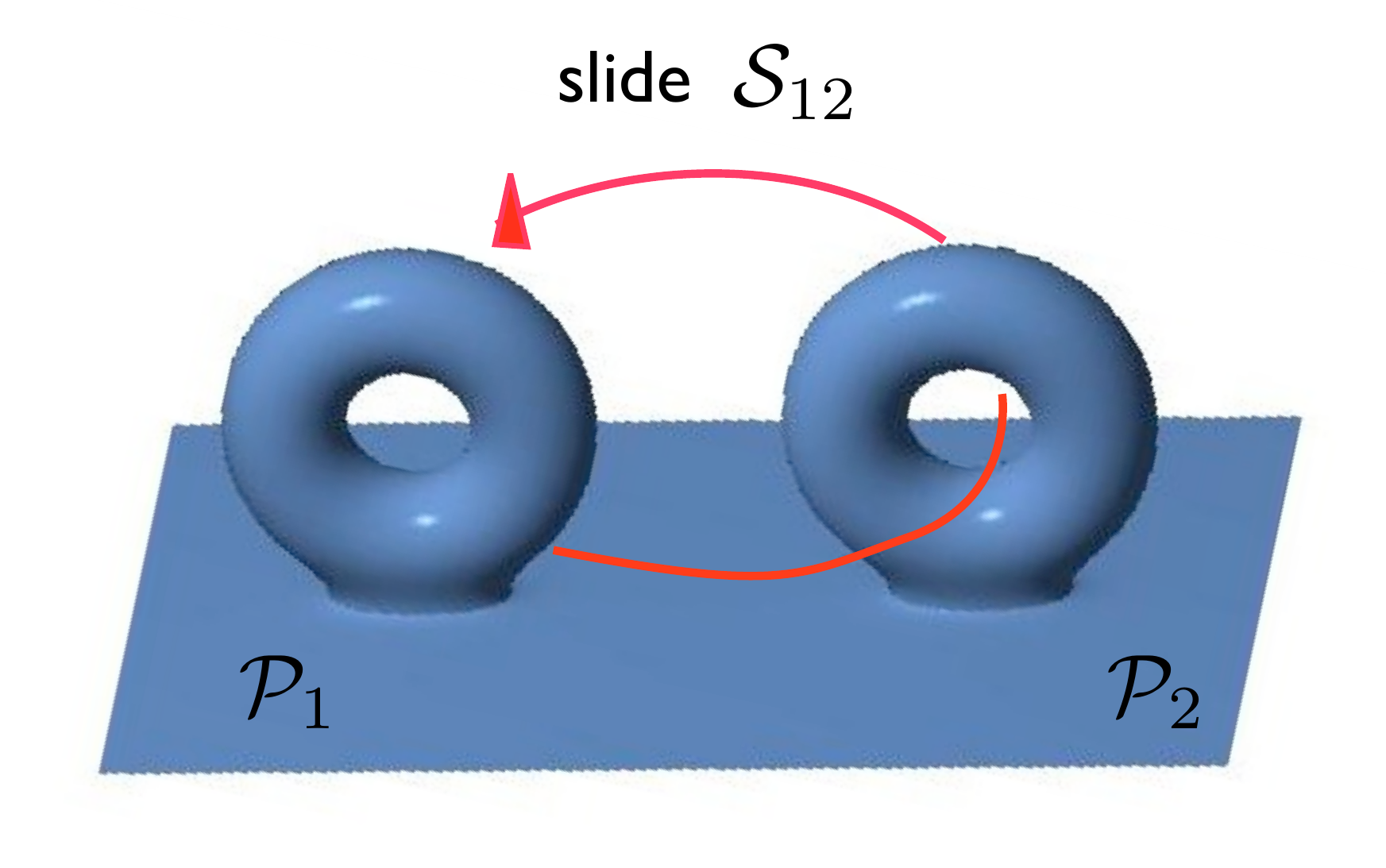} 
\caption{$\R^d\#\mathcal{P}_1\#\mathcal{P}_2$}
\label{fig4}
\end{figure}

If $\R^d\#\pr\#\pr$ is represented as in Fig. \ref{fig4} with bumps representing $\pr$,
the exchange diffeo $E^{(2)}$ can be regarded as moving the geons so that they exchange places. This diffeo (mod $D^{(2)\infty}_0$) is the generator of $S_2$ in (\ref{geo36}). For $d=3$, $E^{(2)2}\in D^{(2)\infty}_0$, but for $d=2$ that is not so. Thus for $d=2$, we can have geons with fractional statistics \cite{Anez}.

Slides $\mathscr{S}^{(2)}$ arise because for $\pr\neq S^d$, $\pi_1(\pr)\neq\{e\}$ for $d=2$ (where $\pr=T^2$), and $d=3$ (in view of the now-proved Poincar\'e conjecture). Thus let $L$ be a non-contractable loop threading $\R^d\#\pr^{(1)}$, where $\pr^{(j)}$ are primes and let $\B_p$ be a ball containing a point $P$ on $L$ in its interior. Then $\mathscr{S}^{(2)}_{21}$, the slide of $\pr^{(2)}$ along $L$ through $\pr^{(1)}$, is obtained by attaching $\pr^{(2)}$ to $\partial\B_p$ and dragging it along $L$ by moving $p$ in a loop around $L$. Note that the slide $\mathscr{S}^{(2)}_{12}$ of $\pr^{(1)}$ through $\pr^{(2)}$ is not equal to $\mathscr{S}_{21}^{(2)}$.

\section{Quantum Fields}

In standard quantum physics, there is a relation between spacetime symmetries like the Poincar\'e group $\Pg$ and the statistics group that implements the identity of particles. It can be described as follows. An element $\alpha$ of the Poincar\'e group acts on a member $\psi$ of the single particle Hilbert space $\mathcal{H}$ by pullback:
\beq
\big(\alpha\psi\big)(x)=\psi(\alpha^{-1}x)\quad.
\eeq
This action extends to the $N$-particle Hilbert space $\mathcal{H}^{\otimes N}$ via the coproduct $\Delta_0$:
\beq\label{geo2}
\Delta_0(\alpha)=\alpha\otimes\alpha
\eeq
Thus on $\mathcal{H}^{\otimes N}$, it acts by
\beq
\underbrace{(\I\otimes\I\otimes...\otimes\Delta_0)}_{(N-1) \ factors}\underbrace{(\I\otimes\I\otimes...\otimes\Delta_0)}_{(N-2) \ factors}...\Delta_0(\alpha)=\underbrace{\alpha\otimes\alpha\otimes...\otimes\alpha}_{N\ factors}
\eeq

The statistics group expressing the identity of particles must commute with the action of the symmetry group. This requirement just says that symmetry transformations, such as Lorentz transformations, should not spoil particle identity. It is fulfilled by the permutation group $S_N$ which permutes the factors in the tensor product
\beq
\psi_1\otimes\psi_2\otimes...\otimes\psi_N\in\mathcal{H}^{\otimes N}
\eeq

Quantum fields compatible with the symmetry group such as $\Pg$ and implementing statistics exist. For these fields, the permutation group $S_N$ and say the Poincar\'e transformation commute when acting on $N$-particle in- or out- states.

In the case of geon spacetimes the situation is more involved. In fact while for the standard coproduct like that in (\ref{geo2}), the internal symmetry $D^{(1)\infty}/D^{(1)\infty}_0$ acts by its diagonal map into $D^{(N)\infty}/D^{(N)\infty}_0$ and that action commutes with $S_N$, the slides present a more complex story. They do not commute with $S_N$ (nor with $\alpha\otimes...\otimes\alpha$ for $\alpha\in D^{(1)\infty}/D^{(1)\infty}_0$) and can change representations of $S_N$: they can convert bosons into fermions! For such reasons, Sorkin and Surya have suggested that elements of $\mathscr{S}$ represent interactions of geons. Elsewhere we will implement this idea in a qft. But for now we let $\mathscr{S}$ act by the identity representation on quantum states. That means that we will work with $\left[\times^ND^{(1)\infty}/D^{(1)\infty}_0\right]\rtimes S_N$ and their group algebra.

The generalisations of symmetry groups are Hopf algebras. This more general setting is needed by us below where we will work with the group algebra $\Da$ with a twisted coproduct. In that case too, the algebra defining statistics is in the commutant of the coproduct. It is still $S_N$, but acts differently on $\mathcal{H}^{\otimes N}$.

From (\ref{geo36}), slides form an invariant subgroup in $D^{(N)\infty}/D^{(N)\infty}_0$. For this reason, slides can be represented by identity on quantum states. Sorkin and Surya \cite{MCG1,MCG2}  have suggested that we do so motivated by the considerations above. We follow their suggestion.

\section{Twists of Geon Spacetimes: The choice}

The Drinfel'd twist $F_\theta$ of the Poincar\'e group algebra $\mathbb{C}\Pg$, which let the latter act on the Moyal plane $\A(\R^d)$, is by now well-known \cite{review}. If $P_\mu$ are the translations generators of $\Pg$ and $\theta=[\theta_{\mu\nu}=-\theta_{\nu\mu}\in\R]$ is the matrix characterising the Moyal $\star$-product,
\bea\label{geo33}
f\star g=f\e^{-\frac{i}{2}\overleftarrow{\mathcal{P}_\mu}\theta^{\mu\nu}\overrightarrow{\mathcal{P}}_\nu}g,\quad f,g\in\A(\R^d),\eea
where $\mathcal{P}_\mu =-i\partial_\mu$ is the representative of  $P_\mu$ on $\A(\R^d)$, then
\beq\label{geo34}
F_\theta=\e^{-\frac{i}{2}P_\mu\otimes\theta^{\mu\nu}P_\nu}\, .
\eeq
Its realization on the algebra of functions is just the term appearing in between the two functions $f$ and $g$ in (\ref{geo33}). We will indicate it by a script $\F$:
\beq\label{geo35}
\F=\e^{\frac{i}{2}\overleftarrow{\partial}_\mu\theta_{\mu\nu}\overrightarrow{\partial}_\nu}\quad.
\eeq

The twisted coproduct $\Delta_\theta$ of $\mathbb{C}\Pg$ is fixed by those of the elements $g\in\Pg$:
\beq\label{geo7}
\Delta_\theta(g)=F^{-1}_\theta(g\otimes g)F_\theta\quad.
\eeq
We want to generalise $F_\theta$ to geonic diffeos and in particular to $D^{(1)\infty}/D^{(1)\infty}_0$. The motivation is as follows.

If a sphere $S^{d-1}$ encloses the prime in $\R^d\#\pr$ in the sense that the complement of this sphere in $\R^d\#\pr$ is homeomorphic to $\R^d/\B^d$ where $\B^d$ is the $d$--dimensional ball, then by suitably adjoining elements of $D^\infty_0$, we can ensure that $D^\infty/D^\infty_0$ acts as the identity outside $S^{d-1}$. So these diffeos can be taken to be localised on the geon. If the geon size is of the order of the Planck volume, the action of $D^{(1)\infty}/D^{(1)\infty}_0$ is also confined to such Planck volumes (It is possible to define geon sizes in a diffeo-invariant way \cite{Anez}). As explained in the introduction, at these scales we expect the spacetime to be noncommutative and the action of the symmetry group to be consequently twisted.

We will generalise $F_\theta$ to $D^{(1)\infty}/D^{(1)\infty}_0$ and after that twist using elements of $D^{(1)\infty}/D^{(1)\infty}_0$. Then, as we shall see, spacetimes become noncommutative on the above Planck-scale volumes. This is in accordance with the arguments of DFR \cite{Doplicher}.

{\it Thus the choice of twists using $D^{(1)\infty}/D^{(1)\infty}_0$ appears to be one good way to implement the DFR ideas}.

It is also one way to incorporate aspects of the topology of geons in these basic quantum field theories as we shall see.

\section{Twists of Geon Spacetimes: Coassociative Coproducts}

The generalisation of $F_\theta$ to $D^{(1)\infty}/D^{(1)\infty}_0$ is not immediate since $D^{(1)\infty}/D^{(1)\infty}_0$ is discrete. It can be finite or infinite, but it is certainly discrete. So we must know how to adapt $F_\theta$ to discrete groups. The difficulty comes from the fact that for Lie groups, we write $F_\theta$ in terms of the exponential of the tensor product of Lie algebra elements, as in (\ref{geo34}). There is no analogue of the Lie algebra for discrete groups. As we will shortly see, writing the twist $F_\theta$ in momentum space sheds light on the path to follow for the generalization.

The plane waves $\e_p$, $\e_p(x)=\e^{ip\cdot x}$, carry the irreducible representations of the translation subgroup of $\mathbb{C}\Pg$. Since
\beq
\mathcal{P}_\mu\e_p=p_\mu\e_p\quad,
\eeq
the restriction of $\mathscr{F}_\theta$ (\ref{geo35}) to $\e_p\otimes\e_q$ is given by
\beq
\mathscr{F}_\theta\e_p\otimes\e_q=\e^{-\frac{i}{2}p_\mu\theta^{\mu\nu}q_\nu}\e_p\otimes\e_q\ .
\eeq

Let $\mathfrak{P}_p$ be the projection operator which acting on functions of $\R^d$ projects to the subspace spanned by $\e_p$. It is thus the projector to the irreducible representation of the translation subgroup identified by the real vector ``$p$''. For a particle of mass $m$, for which $p_0=\sqrt{\vec{p}^2+m^2}$, we can define $\PR_p$ by requiring that
\beq
\PR_p\e_q=2|p_0|\delta^{(3)}(\vec{p}-\vec{q})\e_p\quad.
\eeq

Then we can see that
\beq\label{geo3}
\mathscr{F}_\theta=\int\dx\mu(p)\dx\mu(q)\e^{-\frac{i}{2}p\wedge q}\PR_p\otimes\PR_q,\qquad\dx\mu(p):=\frac{\dx^3p}{2\sqrt{\vec{p}^2+m^2}}
\eeq
where $p\wedge q:=p_\mu\theta_{\mu\nu}q_\nu$, and that
\beq\label{geo4}
F_\theta=\int\dx\mu(p)\dx\mu(q)\e^{-\frac{i}{2}p\wedge q}\PR_p\otimes\PR_q\quad.
\eeq

If $\e_p$ is off--shell, so that $p_0$ is not constrained to be $\sqrt{\vec{p}^2+m^2}$, we can still write $\F$ in terms of projections by slightly modifying (\ref{geo3}).

\subsection{A Simple Generalisation to Discrete Abelian Groups}

It is possible to find a simple generalisation of (\ref{geo3}-\ref{geo4}) to discrete abelian groups. We first discuss this generalisation.

Consider first the group
\beq
\Z_n=\{\xi^k\equiv\e^{i\frac{2\pi}{n}k}:k=0,1,...,(n-1)\}\quad.
\eeq
Its IRR's $\varrho_m$ are all one-dimensional and given by its characters $\chi_m$:
\beq\label{geo6}
\chi_m(\xi)=\xi^m,\quad m\in\{0,1,...,(n-1)\}\quad.
\eeq
Then if $\hat{\xi}$ is the operator representing $\xi$ on the space on which it acts, the projector $\PR_m$ to the IRR $\varrho_m$ is
\beq
\PR_m=\frac{1}{n}\sum^{n-1}_{k=0}\bar{\chi}_m(\xi^k)\hat{\xi}^k
\eeq
This follows from 
\bea
\hat{\xi}^l\PR_m = \frac{1}{n}\sum^{n-1}_{k=0}\bar{\chi}_m(\xi^k)\hat{\xi}^{k+l}
= \frac{1}{n}\sum^{n+l-1}_{k=l}\bar{\chi}_m(\xi^{k-l})\hat{\xi}^k = \chi_m(\xi^l)\PR_m
\eea
where we used the fact that
\beq
\bar{\chi}(\xi^l)\chi(\xi^l)=1,\qquad \bar{\chi}(\xi^l)=\chi(\xi^{-l}),
\eeq
and the orthogonality relations,
\beq
\frac{1}{n}\sum_\xi\bar{\chi}_m(\xi)\chi_n(\xi)=\delta_{m,n} \, ,
\eeq
that imply,
\beq
\PR_m\PR_n=\delta_{m,n}\PR_n\, .
\eeq

Note that $\PR_m$ is the image of
\beq
\mathbb{P}_m=\frac{1}{n}\sum^{n-1}_{k=0}\bar{\chi}_m(\xi^k)\xi^k
\eeq
in the group algebra $\mathbb{C}\Z_n$ and that 
\beq
\mathbb{P}_m\mathbb{P}_n=\delta_{m,n}\mathbb{P}_n,\qquad\sum^{n-1}_{m=0}\mathbb{P}_m=\I\quad.
\eeq

\subsection{The case of $D^{(1)\infty}/D^{(1)\infty}_0$}
From $D^{(1)\infty}/D^{(1)\infty}_0$, we pick its maximal abelian subgroup $A$ and assume for the moment that $A$ is finite. Then $A$ is the direct product of cyclic groups:
\beq
A=\Z_n\times\Z_{n_2}\times...\times\Z_{n_k}\quad.
\eeq
Its IRR's are  given by:
\beq
\varrho_{m_1}\otimes\varrho_{m_2}\otimes...\otimes\varrho_{m_k},\quad m_j\in\{0,1,..,n_j-1\} \, ,
\eeq
with characters
\beq
\chi_{\vec{m}}=\prod_i\chi_{m_i}
\eeq
and projectors $\PR_{\vec{m}}=\otimes_i\PR_{m_i}$ on the representation space or projectors
\beq\label{geo5}
\mathbb{P}_{\vec{m}}=\otimes_i\mathbb{P}_{m_i}, \quad\mathbb{P}_{\vec{m}}\mathbb{P}_{\vec{m'}}=\delta_{\vec{m},\vec{m}'}\mathbb{P}_{\vec{m}},\quad\sum_{\vec{m}}\mathbb{P}_{\vec{m}}={\rm identity\ of}\ A
\eeq
in the group algebra $\mathbb{C}A$. (The summation of $m_j$ in (\ref{geo5}) is from 0 to $n_j-1$).

Let $\theta=[\theta_{ij}=-\theta_{ji}\in \R]$ be an antisymmetric matrix with constant entries. Following (\ref{geo4}), we can write a Drinfel'd twist using elements of $\mathbb{C}A$:
\beq\label{geo9}
F_\theta=\sum_{\vec{m},\vec{m}'}\e^{-{\frac{i}{2}m_i\theta_{ij}m'_j}}\mathbb{P}_{\vec{m}}\otimes\mathbb{P}_{\vec{m}'}\quad.
\eeq

But there are quantisation conditions on $\theta_{ij}$. That is because $\varrho_m$ and $\varrho_{m+n}$ give the same IRR for $\Z_n$ as (\ref{geo6}) shows. That means that $\vec{m}$ and $\vec{m}+(0,...,0,n_i,0,...0)$ give the same IRR $\varrho_{\vec{m}}$, $n_i$ being the $i^{th}$ entry. Since $F_\theta$ must be invariant under these shifts, we find that $\theta_{ij}$ is restricted to the values
\beq
\theta_{ij}=\frac{4\pi}{n_{ij}}
\eeq
where
\beq\label{geo20}
\frac{n_i}{n_{ij}},\ \frac{n_j}{n_{ij}}\in\Z\quad.
\eeq

The twist (\ref{geo7}) of the canonical coproduct of $\Pg$ using $F_\theta$ leads to a coassociative coproduct. Similarly the twist of the coproduct of $D^{(1)\infty}/D^{(1)\infty}_0$ or any of its subgroups leads to a coassociative coproduct. That is because the twist involves the abelian algebra $\mathbb{C}A$. As we will further discuss later on, the spacetime algebra is associative, but not commutative if a $\theta_{ij}=-\theta_{ji}\neq0$ \cite{Mario2}.

\subsection*{{\it Remarks}}
\begin{itemize}
\item[a)] The condition (\ref{geo20}) has a solution $n_{ij}\neq\pm1$ only if $n_i$ and $n_j$ have a common factor ($\neq\pm 1$). Thus if say $n_i=2$, $n_j=3$ for some $i,j$ then $n_{ij}=\pm1$. For either of these solutions,
\beq
\e^{-\frac{i}{2}m_i\theta_{ij}m'_j}=1\quad {\rm or}\quad \theta_{ij}\ {\rm is\ effectively\ equivalent\ to\ 0}\quad.
\eeq

\item[b)] There are many instances where $A$ contains factors of $\Z$. The IRR's $\varrho_\varphi$ of $\Z$ are given by points of $S^1=\{\e^{i2\pi\varphi}:0\leq\varphi\leq1\}$:
\beq
\varrho_\varphi:n\in\Z\to\e^{i2\pi n\varphi}.
\eeq

Note that 
\beq
\varrho_\varphi=\varrho_{\varphi+1}\quad.
\eeq

Suppose now that $A=\times_{i=1}^k\Z_{n_i}\times\Z$. Now its IRR's are labelled by the vector $(\vec{m},\varphi)=(m_1,...,m_k,\varphi)$. The twist $F_\theta$ is written as
\begin{eqnarray}\label{geo8}
F_\theta = \sum_{\vec{m},\vec{m}'} & & \int^1_0\dx\varphi\int^1_0\dx\varphi'\ \e^{-\frac{i}{2}m_i\theta_{ij}m'_j} \times \nonumber \\ & & \times  \e^{-\frac{i}{2}[m_i(\theta_{m_i,k+1})\varphi'-\varphi(\theta_{m_i,k+1})m_i']}\mathbb{P}_{(\vec{m},\varphi)}\otimes\mathbb{P}_{(\vec{m}',\varphi')}
\end{eqnarray}
But the periodicity in $\varphi,\ \varphi'$ is 1 and hence $\theta_{m_i,k+1}=\pm4\pi$ and the second exponential in (\ref{geo8}) is $\I\otimes\I$. In short, $F_\theta$ has no twist factor involving $\Z$ and $F_\theta$ reduces back to the earlier expression (\ref{geo9}). If there are say {\it two factors} of $\Z$ so that $A=\times_{i=1}^{k-1}\Z_{n_i}\otimes\Z\otimes\Z$ the second exponential in (\ref{geo8}) is replaced by
\beq
\e^{\varphi\theta_{k,k+1}\varphi'}
\eeq
and we require its periodicity in $\varphi$ and $\varphi'$. Hence $\theta_{k,k+1}\simeq0$.
In this way, we see that $F_\theta$ depends nontrivially only on compact abelian discrete groups.

\item[c)] Later in section IX, we will argue that the twists found above seem general so long as we insist on the coassociativity of the coproduct (or equivalently the associativity of the spacetime algebra).

\end{itemize}

\section{On Twisted Symmetrisation and Antisymmetrisation}

Let $\mathcal{H}$ be a one-geon Hilbert space. It carries a representation of $D^{(1)}$ or more generally of $D$. The ``momentum constraint'' is implemented by requiring that $D^{\infty}_0\to\I$ in this representation which we assume is satisfied. 

Let $\tau_0$ be the flip operator on $\mathcal{H}\otimes\mathcal{H}$:
\beq
\tau_0\ \alpha\otimes\beta=\beta\otimes\alpha,\quad\alpha,\beta\in\mathcal{H}\quad.
\eeq
When the coproduct is $\Delta_0$, $\Delta_0(d)=d\otimes d$ for $d \in D$,  $\tau_0$ commutes with $\Delta_0$ (by $d$ here we mean the representation of $d$ on $\mathcal{H}$.). So the subspaces $\frac{\I\pm\tau_0}{2}\mathcal{H}\otimes\mathcal{H}$ are invariant under diffeos and carry the identity representation of $D^{\infty}_0$. We can then use them to define bosonic and fermionic geons.

But if we deform $\Delta_0$ into (\ref{geo7}), $\tau_0$ does not commute with $\Delta_\theta(d)$ for all $d$ anymore if $F_\theta\neq\I\otimes\I$. So the subspaces $\frac{\I\pm\tau_0}{2}\mathcal{H}\otimes\mathcal{H}$ are not diffeomorphism invariant, nor need they fulfill the constraint $\Delta_\theta(d)\Big[\tau_0(\alpha\otimes\beta)\Big]=\tau_0\Big[\Delta_\theta(d)(\alpha\otimes\beta)\Big]$ for $d\in D^{(1)}$. That means that bosons and fermions cannot be associated with the subspaces $\frac{\I\pm\tau_0}{2}\mathcal{H}\otimes\mathcal{H}$.

Instead, as discussed elsewhere \cite{mangano,sasha}, one should use the twisted flip operator
\beq
\tau_\theta=F_\theta^{-1}\tau_0F_\theta,\quad\tau_\theta^2=\I\otimes\I
\eeq
which commutes with the twisted coproduct $\Delta_\theta(d)$. Bosonic and fermionic geons are thus associated with the subspaces $\frac{\I\pm\tau_\theta}{2}\mathcal{H}\otimes\mathcal{H}$.

The twist depends on $D^{\infty}/D^{\infty}_0$. So these twisted subspaces incorporate at least aspects of the internal diffeos of geons unlike $\tau_0$. Such a twist of flip is a consequence of deforming the coproduct to $\Delta_\theta$. As we will discuss, this deformation introduces spacetime noncommutativity localised at the geon. Further there are outlines available for an approach to build an orderly quantum field theory (compatibly with the DFR suggestion) incorporating this noncommutativity and deformed statistics, and transforming by the twisted coproduct. These are all attractive aspects of introducing the twist $F_\theta$.

\section{Covariant Quantum Fields: The Moyal Plane}

Elsewhere \cite{covariant}, we have carefully discussed the notion of covariant fields in general including in particular the Moyal plane $\A(\R^d)$. This concept in the limit $\theta\to0$ reduces to the corresponding well-known concept for $\theta=0$. We give a short review of our earlier work \cite{covariant} for the Moyal plane and then adapt it to discrete Hopf algebras

\subsection{The $\theta=0$, Commutative Plane $\mathcal{A}_0(\R^d)$}

Consider a scalar quantum field $\varphi_0$ based on the commutative spacetime algebra $\mathcal{A}_0(\R^d)$. $\varphi_0$ depends on the spacetime point $x$ while $\varphi_0(x)$ is an ``operator'' on a Hilbert space. Each element $g$ of an appropriate diffeo group acts on $x$,
\beq
g:x\to gx
\eeq
and on fields on $\R^d$ by pull-back:
\beq
\varphi_0\to g\varphi_0,\quad (g\varphi_0)(x)=\varphi_0(g^{-1}x)\quad.
\eeq
It acts on $\varphi_0(x)$ also by a unitary operator $U(g)$:
\beq
\varphi_0(x)\to U(g)\varphi_0(x)U(g)^{-1}\quad.
\eeq

The field is said to be {\it covariant} if it is invariant under the combined action of both:
\beq
U(g)\varphi_0(g^{-1}x)U(g)^{-1}=\varphi_0(x)\quad.
\eeq

In the literature, this equation is often written as
\beq
U(g)\varphi_0(x)U(g)^{-1}=\varphi_0(gx)\quad.
\eeq

Poincar\'e invariant qft's are based on covariant fields. Covariance ensures that the spacetime action of the Poincar\'e group is implementable by unitary operators. Its classical version would ensure that spacetime transformations are canonically implemented.

A Poincar\'e-invariant qft has a unique vacuum $|0\rangle$ invariant under $U(g):U(g)|0\rangle=|0\rangle$. It is a cyclic vector. Using this fact, we can rewrite the covariance of fields in the following way:
\beq\label{geo10}
U(g)\varphi_0(g^{-1}x_1)\varphi_0(g^{-1}x_2)...\varphi_0(g^{-1}x_n)|0\rangle=\varphi_0(x_1)\varphi_0(x_2)...\varphi_0(x_n)|0\rangle\quad.
\eeq
But note that this uses the canonical coproduct $\Delta_0(g)$ when transforming the spacetime arguments.

The implications of this equation become transparent if written in terms of in (or out) fields. The in field $\varphi^{{\rm in}}_0$ of mass $m$ has the expansion
\beq\label{geo10b}
\varphi^{{\rm in}}_0=\int\dx\mu(p)\Big[c^{{\rm in}\dag}_p\e_p+h.c.\Big],\quad \e_p=\e^{-ip\cdot x},\quad\dx\mu(p)=\frac{\dx^3p}{2\sqrt{\vec{p}^2+m^2}}\quad.
\eeq
We consider the implications of (\ref{geo10}) using (\ref{geo10b}) for translations and Lorentz transformations in turn.

\subsubsection{{\it Translations}}

Let first check for covariance under translation.

If $P_\mu$ are the translation generators on the Hilbert space with
\beq\label{geo27}
[P_\mu,c^{{\rm in}\dag}_p]=p_\mu c^{{\rm in}\dag}_p
\eeq
and $\mathcal{P}_\mu=-i\partial_\mu$, the representative of translation generators on spacetime, we have
\beq\label{geo11}
[P_\mu,c^{{\rm in}\dag}_p]\e_p+c^{{\rm in}\dag}_p(\mathcal{P}_\mu\e_p)=0\quad.
\eeq

So $\varphi_0$ has translational covariance.  Notice that we can deduce  (\ref{geo27}) from (\ref{geo11}) and $P_\mu \e_p = -p_\mu \e_p$. 

We can now check (\ref{geo10}) for multi-particle states in terms of creation-annihilation operators. The coproduct $\Delta_0(g)=g\otimes g$ gives
\beq
\Delta_0(\mathcal{P}_\mu)=\mathcal{P}_\mu\otimes\I+\I\otimes\mathcal{P}_\mu\quad.
\eeq
From this follows that (\ref{geo10}) is satisfied at the two-particle level as well. Explicitly,
\begin{eqnarray}
\qquad\qquad\int\prod_i\dx\mu(p_i)\Big[P_\mu,c^{{\rm in}\dag}_{p_1}c^{{\rm in}\dag}_{p_2}\Big]|0\rangle\e_{p_1}\otimes\e_{p_2} + \hskip 3cm &&  \nonumber \\ + \int\prod_i\dx\mu(p_i)c^{{\rm in}\dag}_{p_1}c^{{\rm in}\dag}_{p_2}|0\rangle\Delta_0(\mathcal{P}_\mu)\e_{p_1}\otimes\e_{p_2} &&   =   \nonumber \\
 = \int\prod_i\dx\mu(p_i)\left\{\left(\Big[P_\mu,c^{{\rm in}\dag}_{p_1}\Big]c^{{\rm in}\dag}_{p_2}+c^{{\rm in}\dag}_{p_1}\Big[P_\mu,c^{{\rm in}\dag}_{p_2}\Big]\right)|0\rangle\e_{p_1}\otimes\e_{p_2} \right. && + \nonumber \\ + \left. c^{{\rm in}\dag}_{p_1}c^{{\rm in}\dag}_{p_2}|0\rangle\Big[\mathcal{P}_\mu\otimes\I+\I\otimes\mathcal{P}_\mu\Big]\e_{p_1}\otimes\e_{p_2}\right\}   = 0 \, . \hskip .1cm &&
\end{eqnarray}
From this calculation, it is evident that (\ref{geo10}) generalises to generic $N$:
\begin{eqnarray}
&&\int\prod_i\dx\mu(p_i)\Big[P_\mu,c^{{\rm in}\dag}_{p_1}...c^{{\rm in}\dag}_{p_N}\Big]|0\rangle\e_{p_1}\otimes...\otimes\e_{p_N}+ \\
&& + \int\prod_i\dx\mu(p_i)c^{{\rm in}\dag}_{p_1}...c^{{\rm in}\dag}_{p_N}|0\rangle\Big[\I\otimes\I\otimes...\otimes\Delta_0\Big]...\Big[\I\otimes\Delta_0\Big]\Delta_0(\mathcal{P}_\mu)\e_{p_1}\otimes...\otimes\e_{p_N}\,.  \nonumber
\end{eqnarray}
Thus (\ref{geo11}) is necessary for covariance.

The consistency of the adjoint of (\ref{geo27}),
\beq
[P_\mu,c^{{\rm in}}_p]=-p_\mu c^{{\rm in}}_p
\eeq
with covariance can also be established in the same way by acting with fields on bra vectors.

\subsubsection{{\it Lorentz Transformations}}

Next we consider covariance under Lorentz transformations $\Lambda\in\Pg$ and the transformation rules it implies on $c^{{\rm in}\dag}_p$, $c^{{\rm in}}_p$. Under $\Lambda$,
\beq
\Lambda:\e_p\to \e_{\Lambda p}
\eeq
while $\dx\mu(\Lambda p)=\dx\mu(p)$. Thus the equation
\beq
U(\Lambda)c^{{\rm in}\dag}_pU(\Lambda)^{-1}=c^{{\rm in}\dag}_{\Lambda p}
\eeq
is immediate, while its adjoint is
\beq
U(\Lambda)c^{{\rm in}}_pU(\Lambda)^{-1}=c^{{\rm in}}_{\Lambda p}
\eeq
As before it generalises to products of $N$ $c^{{\rm in}}_p$'s.

\subsection{The Moyal Plane $\A(\R^d)$}

Let us next review the Moyal case. For single particle states, the twisted and untwisted actions do not differ, so we consider the two-particle sector.

The twisted in fields will be written as
\beq
\varphi_\theta^{{\rm in}}=\int\dx\mu(p)\Big[a^{{\rm in}\dag}_p\e_p+h.c.\Big]
\eeq
where the relation of $a^{{\rm in}\dag}_p$, $a^{{\rm in}}_p$ to $c^{{\rm in}\dag}_p$, $c^{{\rm in}}_p$ will be determined from covariance. 

As usual we assume that there is a unique Poincar\'e-invariant vacuum annihilated by $a^{{\rm in}}_p$.

The unitary operators representing $g\in\Pg$ on the Hilbert space of states will be denoted by $U_\theta(g)$ with $U_0(g)=U(g)$. We will show that $U_\theta(g)$ has the same expression as $U(g)$ in terms of $c^{{\rm in}\dag}_p$, $c^{{\rm in}}_p$ so that we will later write
\beq
U_\theta(g)=U(g)\, .
\eeq

Consider
\beq
\int\dx\mu(p_1)\dx\mu(p_2)a^{{\rm in}\dag}_{p_1}a^{{\rm in}\dag}_{p_2}|0\rangle\e_{p_1}\otimes\e_{p_2} \, .
\eeq

The twisted coproduct is\footnote{We want to explain our notation to the reader here. Previously we already introduced the ``definition'' of the wedge symbol among four-vector as
\beq\nonumber
p_1\wedge p_2:=p_{1\mu}\theta^{\mu\nu}p_{1\nu}\, .
\eeq
(This differs from the wedge symbol among differential forms.)
In the following we extend this definition to operators as well. In the case of a bi-operator, a tensor product is to be understood as implicit when a wedge symbol appears,
\beq\nonumber
P\wedge P:=P_{\mu}\theta^{\mu\nu}\otimes P_{\nu}
\eeq
whereas there is no tensor product symbol in the case when one of the $p$'s is a vector:
\beq\nonumber
p\wedge P:=p_{\mu}\theta^{\mu\nu}P_\nu\, .
\eeq
These notations should not cause confusion.
}
\beq
\Delta_\theta(g)=F^{-1}_\theta(g\otimes g)F_\theta,\quad F_\theta=\e^{-\frac{i}{2}P\wedge P}\to \mathscr{F}_\theta=\e^{\frac{i}{2}\overleftarrow{\partial}\wedge\overrightarrow{\partial}}\ {\rm on}\ \e_{p_1}\otimes\e_{p_2} \, .
\eeq

As
\beq
\Delta_\theta(P_\mu)=\Delta_0(P_\mu)=P_\mu\otimes\I+\I\otimes P_\mu\, ,
\eeq
we find as before that
\beq
[P_\mu,a^{{\rm in}\dag}_p]=p_\mu a^{{\rm in}\dag}_p,\quad [P_\mu,a^{{\rm in}}_p]=-p_\mu a^{{\rm in}}_p  \, .
\eeq

As for Lorentz transformations,
\beq\label{Eq1}
\Delta_\theta(\Lambda)\e_{p_1}\otimes\e_{p_2}=\e^{\frac{i}{2}(\Lambda p_1)\wedge(\Lambda p_2)}\e^{-\frac{i}{2}p_1\wedge p_2}\e_{\Lambda p_1}\otimes\e_{\Lambda p_2}  \, .
\eeq

Covariance hence requires that 
\bea\label{geo13}
U_\theta(\Lambda) && \int\dx\mu(p_1)\dx\mu(p_2)a^{{\rm in}\dag}_{p_1}a^{{\rm in}\dag}_{p_2}|0\rangle\e^{\frac{i}{2}(\Lambda p_1)\wedge(\Lambda p_2)}\e^{-\frac{i}{2}p_1\wedge p_2} = \nonumber \\ && = \int\dx\mu(p_1)\dx\mu(p_2)a^{{\rm in}\dag}_{p_1}a^{{\rm in}\dag}_{p_2}|0\rangle\, .
\eea

A solution of this equation is 
\beq\label{geo12}
a^{{\rm in}\dag}_{p}=c^{{\rm in}\dag}_{p}\e^{\frac{i}{2}p\wedge P},\quad U_\theta(\Lambda)=U_0(\Lambda)\quad.
\eeq
We will argue below that this solution is unique.

At the level of wave functions, compatibility with the coproduct $\Delta_\theta$ requires that we twist-symmetrise using
\beq
\tau_\theta=F_\theta^{-1}\tau_0F_\theta,\quad\tau_0\alpha\otimes\beta=\beta\otimes\alpha\quad.
\eeq
If $\otimes_{S_\theta}$ denotes twisted Bose-symmetrised tensor product,
\beq
\e_{p_1}\otimes_{S_\theta}\e_{p_2}=\frac{\I+\tau_\theta}{2}\e_{p_1}\otimes\e_{p_2}\quad,
\eeq
then
\beq
\e_{p_1}\otimes_{S_\theta}\e_{p_2}=\frac{1}{2}\Big[\e_{p_1}\otimes\e_{p_2}+\e^{ip_1\wedge p_2}\e_{p_2}\otimes\e_{p_1}\Big]\quad,
\eeq
and
\beq
\e_{p_2}\otimes_{S_\theta}\e_{p_1}=\frac{1}{2}\Big[\e_{p_2}\otimes\e_{p_1}+\e^{-ip_1\wedge p_2}\e_{p_1}\otimes\e_{p_2}\Big]=\e^{-ip_1\wedge p_2}\e_{p_1}\otimes_{S_\theta}\e_{p_2}\quad.
\eeq

Thus the equality of 
\beq
\int\dx\mu(p_1)\dx\mu(p_2)a^{{\rm in}\dag}_{p_1}a^{{\rm in}\dag}_{p_2}|0\rangle\e_{p_1}\otimes_{S_\theta}\e_{p_2}
\eeq
and
\beq
\int\dx\mu(p_1)\dx\mu(p_2)a^{{\rm in}\dag}_{p_1}a^{{\rm in}\dag}_{p_2}|0\rangle e^{ip_1\wedge p_2} \e_{p_2}\otimes_{S_\theta}\e_{p_1}
\eeq
requires that
\beq
a^{{\rm in}\dag}_{p_2}a^{{\rm in}\dag}_{p_1}|0\rangle=\e^{-ip_1\wedge p_2}a^{{\rm in}\dag}_{p_1}a^{{\rm in}\dag}_{p_2}|0\rangle\quad.
\eeq

A solution is (\ref{geo12}). It extends correctly to similar identities among $N$-fold tensor products. It thus seems unique. We can then take (\ref{geo12}) to be the relation between untwisted and twisted fields.

Suppose now that we transform $a^{{\rm in}\dag}_{p_1}$ and $a^{{\rm in}\dag}_{p_2}$ with $U_0(\Lambda)$:
\beq
U_0(\Lambda)a^{{\rm in}\dag}_{p_1}a^{{\rm in}\dag}_{p_2}U_0(\Lambda)^{-1}|0\rangle=a^{{\rm in}\dag}_{\Lambda p_1}a^{{\rm in}\dag}_{\Lambda p_2}|0\rangle\e^{-\frac{i}{2}(\Lambda p_1)\wedge(\Lambda p_2)}\e^{\frac{i}{2}p_1\wedge p_2}\quad.
\eeq
This is as required by (\ref{geo13}). This relation extends to $N$-particle states. Also since the Moyal $\star$-product is translationally invariant, $U_\theta(a)=U_0(a)$ for translation $a\equiv(a,\I)\in\mathscr{P}$. Hence we conclude that $U_0(g)\equiv U(g)$ implements the Poincar\'e group on the Hilbert space of states for the Moyal plane \cite{sasha}.

\subsection{Self-Reproduction}

If $\alpha,\beta\in C^\infty(\R^d)$ are scalar functions, then so is $\alpha\beta$ on the commutative spacetime $\mathcal{A}_0(\R^d)$.  {\it Self-reproduction} extends this fact to $\mathcal{A}_\theta(\R^d)$.  

The $\star$-product on the Moyal plane is given by
\beq\label{geo15}
\alpha\star\beta=m_0\Big(\F\alpha\otimes\beta\Big)\, , 
\eeq
with $m_0\Big(\alpha\otimes\beta\Big)$ is just the pointwise multiplication map $\alpha\beta$ of  $\alpha$ and $\beta$.
On plane waves
\beq
\e_{p_1}\star\e_{p_2}=\e^{-\frac{i}{2}p_1\wedge p_2}\e_{p_1+p_2}
\eeq

Now consider
\beq\label{geo14}
\varphi^{\rm in}_\theta=\int\dx\mu(p)\Big[a^{{\rm in}\dag}_p\e_p+h.c.\Big]=\varphi_0^{\rm in}\e^{-\frac{1}{2}\overleftarrow{\partial}\wedge P}
\eeq
where we used (\ref{geo12}).

If $\varphi_\theta^{\rm in}\star\varphi_\theta^{\rm in}$ is the $\star$-product of these fields (the $\star$ being used in taking products of plane waves), and $\varphi^{\rm in}_\theta\varphi_\theta^{\rm in}$ denotes their point-wise product, then it follows that
\beq
\varphi_\theta^{\rm in}\star\varphi_\theta^{\rm in}=\Big(\varphi_0^{\rm in}\varphi_0^{\rm in}\Big)\e^{-\frac{1}{2}\overleftarrow{\partial}\wedge P}
\eeq
From this, we can easily prove that $\varphi^{\rm in}_\theta\star\varphi^{\rm in}_\theta$ transforms with the twisted coproduct under $U(g)$ just as $\varphi_\theta^{\rm in}$ does. This extends to $N$-fold $\star$-products of $\varphi^{\rm in}_\theta$.

The products of quantum fields of course have only a formal significance. Instead we can twist the functions $\e_p$ to $\e^\theta_p$ and implement (\ref{geo14}):
\beq
\varphi^{\rm in}_\theta=\int\dx\mu(p)\Big[c^{{\rm in}\dag}_p\e^\theta_p+h.c.\Big],\quad\e^\theta_p=\e^0_p\e^{\frac{i}{2}p\wedge\mathcal{P}},\quad\mathcal{P}_\mu=-i\partial_\mu
\eeq
where $\e^0_p=e_p$.
Then using (\ref{geo15}), we get
\beq
\e_{p_1}^\theta\star\e_{p_2}^\theta=(\e^0_{p_1}\e^0_{p_2})\e^{\frac{i}{2}(p_1+p_2)\wedge\mathcal{P}}
\eeq
or for generic twisted functions $\alpha^\theta$ and $\beta^\theta$,
\beq
\alpha^\theta=\alpha\e^{-\frac{1}{2}\overleftarrow{\partial}\wedge\mathcal{P}},\quad\beta^\theta=\beta\e^{-\frac{1}{2}\overleftarrow{\partial}\wedge\mathcal{P}},\quad\alpha^\theta\star\beta^\theta=(\alpha\beta)\e^{-\frac{1}{2}\overleftarrow{\partial}\wedge\mathcal{P}}
\eeq
showing that this twist preserves the transformation properties of the Moyal algebra.

\section{Covariant Quantum Fields: Geons on Commutative Spacetimes}

We assume that a covariant quantum field $\varphi_0$ can be associated with a geon when the underlying spacetime is commutative. Diffeomorphism invariance implies that $D^{(1)\infty}_0$ acts trivially on $\varphi_0$. So the group $D^{(1)}/D^{(1)\infty}_0$ acts nontrivially on $\varphi_0$ by the pull-back of the action of $D^{(1)}$ on  spacetime: if $g\in D/D^\infty_0$\footnote{From now on we will only refer to the single geon diffeo group. Therefore we will use $D$, $D^\infty$ and $D^\infty_0$ instead of $D^{(1)}$, $D^{(1)\infty}$  and $D^{(1)\infty}_0$ to simplify the notation.} and $\hat{g}=gg^\infty_0,\ g^\infty_0\in D^\infty_0$ is any member of the equivalence class $gD^\infty_0$, then
\beq\label{geo16}
g:\varphi_0\to g\varphi_0,\quad (g\varphi_0)(p)=\varphi_0(\hat{g}^{-1}p)
\eeq
This action does not depend on the choice of $g^\infty_0$ since $g^\infty_0\varphi_0=\varphi_0$ for all $g_0^\infty\in D^\infty_0$, and hence is consistent.

Equation (\ref{geo16}) has been written for scalar geon fields for simplicity. It is easily generalised to spinorial and tensorial fields.

Also for simplicity, we will henceforth write
\beq
(g\varphi_0)(p)=\varphi_0(g^{-1}p)
\eeq
even though on the r.h.s., we should write $\hat{g}^{-1}p$.

Covariance implies that there exists a representation $U$ of $D/D^\infty_0$ so that
\beq
U(g)\varphi_0(g^{-1}p)U(g)^{-1}=\varphi_0(p)
\eeq
or
\beq
U(g)\varphi_0(p)U(g)^{-1}=\varphi_0(gp)\quad.
\eeq

The twist we now consider is based on abelian discrete compact groups $A$: as we saw, its dependence on representations of $\Z$ is trivial. Let $f^{(\pm)}_{\vec{m}}$ furnish the orthonormal basis on the geon spacetime which carry the UIRR $\vec{m} = (m_1, m_2, \ldots, m_k)$ of $A=\times^k_{i=1}\Z_{n_i}$ and which have positive and negative frequencies $\pm|E_{\vec{m}}|$\footnote{We assume their existence as is normally the case.}:
\bea
f^{(\pm)}_{\vec{m}}(h^{-1}p)&=&f^{(\pm)}_{\vec{m}}(p)\chi_{\vec{m}}(h),\quad h\in A\quad,\\
i\partial_0f^{(\pm)}_{\vec{m}}&=&\pm|E_{\vec{m}}|f^{(\pm)}_{\vec{m}}
\eea
Here $\chi_{\vec{m}}$ is the character function of $A$. Since $\bar{\chi}_{\vec{m}}=\chi_{-\vec{m}}$, we can assume that
\beq
\bar{f}_{\vec{m}}^{(\pm)}=f^{(\mp)}_{-\vec{m}}
\eeq
If $g\in D/D^\infty_0$, we can then write
\beq
f^{(\pm)}_{\vec{m}}(g^{-1}p)=\sum_{\vec{m}'}f^{(\pm)}_{\vec{m}'}(p)\mathscr{D}_{\vec{m}'\vec{m}}(g)
\eeq
where $\mathscr{D}$ is a unitary representation of $D/D^\infty_0$.

The untwisted quantum field $\varphi_0$, assumed real for simplicity, and also assumed to be in, out or free field, can be written as
\beq
\varphi_0=\sum_{\vec{m}}\Big[c_{\vec{m}}f^{(+)}_{\vec{m}}+c^\dag_{\vec{m}}f^{(-)}_{-\vec{m}}\Big]
\eeq

Here $c_{\vec{m}}$, $c^\dag_{\vec{m}}$ are annihilation and creation operators:
\bea\label{geo17}
&[c_{\vec{m}},c^\dag_{\vec{n}}]=\delta_{\vec{m},\vec{n}}\quad,&\\\label{geo28}
\ &[c_{\vec{m}},c_{\vec{n}}]=[c^\dag_{\vec{m}},c^\dag_{\vec{n}}]=0\quad.&
\eea

Covariance is the requirement that there is a unitary representation of $D/D^\infty_0$ on the Hilbert space of vector states such that
\beq
U(g)\varphi_0(g^{-1}p)U(g)^{-1}=\varphi_0(p)\quad.
\eeq
Hence since $\bar{\mathscr{D}}_{\vec{m}'\vec{m}}(g)\mathscr{D}_{\vec{n}'\vec{m}}(g)=\delta_{\vec{m}',\vec{n}'}$ (with sum over $\vec{m}$ being implicit),
\bea\label{geo18}
U(g)c_{\vec{m}}U(g)^{-1}=c_{\vec{m}'}\bar{\mathscr{D}}_{\vec{m}'\vec{m}}(g)\quad,\\\label{geo29}
U(g)c^\dag_{\vec{m}}U(g)^{-1}=c^\dag_{\vec{m}'}\mathscr{D}_{\vec{m}'\vec{m}}(g)\quad.
\eea

For untwisted fields, the symmetrisation postulates on $f^{(\pm)}_{\vec{m}}$ are based on Bose statistics for tensorial fields. They are incorporated in (\ref{geo17},\ref{geo28}) and are compatible with covariance.

\subsection{Covariance for Abelian Twists}

The twisted quantum field $\varphi_\theta$ associated with $\varphi_0$ is written as
\beq
\varphi_\theta=\sum_{\vec{m}}\Big[a_{\vec{m}}f^{(+)}_{\vec{m}}+a^\dag_{\vec{m}}f^{(-)}_{-\vec{m}}\Big]
\eeq
We will as before deduce the relation of $a_{\vec{m}}$, $a^\dag_{\vec{m}}$ to $c_{\vec{m}}$, $c^\dag_{\vec{m}}$ using covariance.

First consider
\beq
\varphi_\theta^{(-)}|0\rangle=\sum_{\vec{m}}a^\dag_{\vec{m}}|0\rangle f^{(-)}_{-\vec{m}}=\sum_{\vec{m}}a^\dag_{\vec{m}}|0\rangle\bar{f}^{(+)}_{\vec{m}}\quad.
\eeq
Covariance implies the requirement
\beq
\sum_{\vec{m}}U(g)a^\dag_{\vec{m}} U(g)^{-1}|0\rangle\bar{f}^{(+)}_{\vec{m}'}\bar{\mathscr{D}}_{\vec{m}'\vec{m}}(g)=\varphi^{(-)}_\theta|0\rangle
\eeq
where $U(g)$ represents $g$ on the vector states and we use $U(g)|0\rangle=|0\rangle$.
Hence 
\beq
U(g)a^\dag_{\vec{m}}U(g)^{-1}|0\rangle=\sum_{\vec{m}''}a^\dag_{\vec{m}''}|0\rangle\mathscr{D}_{\vec{m}''\vec{m}}(g)
\eeq

Next consider the two-particle case:
\beq
\varphi_\theta^{(-)}\otimes\varphi^{(-)}_\theta|0\rangle=\sum_{\vec{m},\vec{n}}a^\dag_{\vec{m}}a^\dag_{\vec{n}}|0\rangle\bar{f}^{(+)}_{\vec{m}}\otimes\bar{f}_{\vec{n}}^{(+)}
\eeq
The action of $g\in D/D^\infty_0$ on $\bar{f}_{\vec{m}}^{(+)}\otimes\bar{f}^{(+)}_{\vec{n}}$ is via the twisted coproduct:
\bea\nonumber
g\triangleright \bar{f}^{(+)}_{\vec{m}}\otimes\bar{f}^{(+)}_{\vec{n}}&=&F_\theta^{-1}(g\otimes g)F_\theta\bar{f}^{(+)}_{\vec{m}}\otimes\bar{f}^{(+)}_{\vec{n}}\\
&=&F^{-1}_\theta(g\otimes g)\bar{f}^{(+)}_{\vec{m}}\otimes\bar{f}^{(+)}_{\vec{n}}\e^{-\frac{i}{2}m_i\theta_{ij}n_j}\\\nonumber
&=&F^{-1}_\theta\sum_{\vec{m}',\vec{n}'}\bar{f}^{(+)}_{\vec{m}'}\otimes\bar{f}^{(+)}_{\vec{n}'}\bar{\mathscr{D}}_{\vec{m}'\vec{m}}(g)\bar{\mathscr{D}}_{\vec{n}'\vec{n}}(g)\e^{-\frac{i}{2}m_i\theta_{ij}n_j}\\\nonumber
&=&\sum_{\vec{m}',\vec{n}'}\bar{f}^{(+)}_{\vec{m}'}\otimes\bar{f}^{(+)}_{\vec{n}'}\e^{\frac{i}{2}m'_i\theta_{ij}n'_j}\bar{\mathscr{D}}_{\vec{m}'\vec{m}}(g)\bar{\mathscr{D}}_{\vec{n}'\vec{n}}(g)\e^{-\frac{i}{2}m_i\theta_{ij}n_j}\quad.
\eea

The covariance requirement
\beq
\sum_{\vec{m},\vec{n}}U(g)a^\dag_{\vec{m}}a^\dag_{\vec{n}}|0\rangle\Big(g\triangleright\bar{f}^{(+)}_{\vec{m}}\otimes\bar{f}^{(+)}_{\vec{n}}\Big)=\varphi^{(-)}_\theta\otimes\varphi_\theta^{(-)}|0\rangle
\eeq
can thus be fulfilled by setting
\beq\label{geo19}
a^\dag_{\vec{m}}=\sum_{\vec{m}'}c^\dag_{\vec{m}}\e^{\frac{i}{2}m_i\theta_{ij}m'_j}\mathbb{P}_{\vec{m}'}
\eeq
and identifying $U(g)$ as the untwisted operator with the action (\ref{geo18},\ref{geo29}) on $c_{\vec{m}}$, $c^\dag_{\vec{m}}$.

The adjoint of (\ref{geo19}) gives
\beq
a_{\vec{m}}=\sum_{\vec{m}'}\Big(\e^{-\frac{i}{2}m_i\theta_{ij}m'_j}\mathbb{P}_{\vec{m}'}\Big)c_{\vec{m}}\equiv V_{-\vec{m}}\ c_{\vec{m}}
\eeq

Now $V_{-\vec{m}}$ is unitary with inverse
\beq
V^{-1}_{-\vec{m}}=V_{\vec{m}}=\sum_{\vec{m}'}\e^{\frac{i}{2}m_i\theta_{ij}m'_j}\mathbb{P}_{\vec{m}'}
\eeq
It is the unitary operator on the quantum HIlbert space representing the element
\beq
\times_j\e^{\frac{i}{2}m_i\theta_{ij}}\in A
\eeq
(The quantisation condition on $\theta_{ij}$ is also manifest from here.) Hence
\beq
V_{\vec{m}}a_{\vec{m}}V^{-1}_{\vec{m}}=\e^{-\frac{i}{2}m_i\theta_{ij}m_j}a_{\vec{m}}=a_{\vec{m}}
\eeq
and 
\beq
a_{\vec{m}}=\sum_{\vec{m}'}c_{\vec{m}}\e^{-\frac{i}{2}m_i\theta_{ij}m'_j}\mathbb{P}_{m'_j}
\eeq
so that we can freely twist on left or right.

The twisted symmetrisation properties (statistics) of the multigeon states
\beq
a^\dag_{\vec{m_1}}a^\dag_{\vec{m_2}}...a^\dag_{\vec{m_N}}|0\rangle
\eeq
follows from (\ref{geo19}).

Self-reproduction under the $\star$-product can also be easily verified:
\beq
\big(a_{\vec{m}}f^{(+)}_{\vec{m}}\big)\star\big(a_{\vec{n}}f^{(+)}_{\vec{n}}\big)=\Big(c_{\vec{m}}c_{\vec{n}}\Big)\e^{\frac{i}{2}(m_i+n_i)\theta_{ij}m'_j}f^{(+)}_{\vec{m}+\vec{n}}\PR_{m'_j}
\eeq
There are similar equation involving creation operators. Here again $\PR_{\vec{m}} = \prod_j\PR_{m_j}$ be the projection operator which acting on functions, projects out the IRR $\vec{m}$ of $A$. With this notation, we can incorporate the dressing transformation directly in $\varphi_\theta$:
\beq
\varphi_\theta=\sum_{\vec{m},\vec{m}'}\Big(\PR_{\vec{m}}\varphi\Big)\e^{-\frac{i}{2}m_i\theta_{ij}m'_j}\mathbb{P}_{\vec{m}'}\quad.
\eeq

This equation is the analogue of the dressing transformation for the Moyal field.

\section{How may we generalise?}

Physical considerations outlined below suggest that the twist discussed above (and its generalisations such as that in the Moyal case) is unique upto unitary equivalence if we require the spacetime algebra to be associative. We do have nonassociative examples \cite{Nonass}, they are associated with quasi-Hopf algebras as symmetries. We will now briefly consider them as well.

\subsection{Abelian Twists $\Rightarrow$ Associative Spacetimes}

For the abelian algebra, we retain $A=\times^k_{i=1}\Z_{n_i}$. If $f^{(\eta)}_{\vec{m}}$, $(\eta=\pm)$, denote the same functions as before, then for the $\star$-product, we assume the general form
\beq
f^{(\eta)}_{\vec{m}}\star f^{(\varrho)}_{\vec{m}'}=\sigma(\vec{m},\vec{m}')f^{(\eta)}_{\vec{m}}f^{(\varrho)}_{\vec{m}'},\quad\eta,\varrho=\pm,\quad\sigma(\vec{m},\vec{m}')\in\mathbb{C}\quad,
\eeq
where on the right, $f^{(\eta)}_{\vec{m}}f^{(\varrho)}_{\vec{m}'}$ denotes point-wise product.

Now $f^{(\eta)}_{\vec{m}}f^{(\varrho)}_{\vec{m}'}$ transforms by the representation $\vec{m}+\vec{m}'$ (modulo $n_i$ in each entry). Taking this into account we require associativity:
\beq
f^{(\eta)}_{\vec{m}}\star\Big(f^{(\varrho)}_{\vec{m}'}\star f^{(\zeta)}_{\vec{m}''}\Big)=\Big(f^{(\eta)}_{\vec{m}}\star f^{(\varrho)}_{\vec{m}'}\Big)\star f^{(\zeta)}_{\vec{m}''}\quad.
\eeq

The l.h.s. and r.h.s. of this equation are
\bea
{\rm l.h.s.}&=&\sigma(\vec{m},\vec{m}'+\vec{m}'')\sigma(\vec{m}',\vec{m}'')f^{(\eta)}_{\vec{m}}f^{(\varrho)}_{\vec{m}'}f^{(\zeta)}_{\vec{m}''}\\
{\rm r.h.s.}&=&\sigma(\vec{m},\vec{m}')\sigma(\vec{m}+\vec{m}',\vec{m}'')f^{(\eta)}_{\vec{m}}f^{(\varrho)}_{\vec{m}'}f^{(\zeta)}_{\vec{m}''}
\eea
Therefore
\beq
\sigma(\vec{m},\vec{m}'+\vec{m}'')\sigma(\vec{m}',\vec{m}'')=\sigma(\vec{m},\vec{m}')\sigma(\vec{m}+\vec{m}',\vec{m}'')\quad.
\eeq
It has the solution
\beq
\sigma(\vec{m},\vec{m}')=\e^{-\frac{i}{2}m_i\hat{\theta}_{ij}m'_j}
\eeq
where $\hat{\theta}_{ij}$ is quantised as before:
\beq
\hat{\theta}_{ij}=\frac{4\pi}{n_{ij}},\quad\frac{n_i}{n_{ij}},\frac{n_j}{n_{ij}}\in\Z\quad.
\eeq

Note that the quantisation requirement forces $\hat{\theta}_{ij}$ to be real, but not necessarily antisymmetric. Hence we can in general write
\beq
\hat{\theta}_{ij}=\theta_{ij}+s_{ij},\quad\theta_{ij}=-\theta_{ij}=\frac{4\pi}{n_{ij}},\quad s_{ij}=s_{ji}=\frac{4\pi}{m_{ij}}
\eeq
where both $n_{ij}$ and $m_{ij}$ divide $n_i$ and $n_j$, that is fulfill the analogue of (\ref{geo20}).

Thus associativity and quantisation conditions reduce $\sigma$ to the form
\beq
\sigma(\vec{m},\vec{m}')=\e^{-\frac{i}{2}m_i\theta_{ij}m'_j}\ \e^{\frac{i}{2}m_is_{ij}m'_j}
\eeq
with the constraints on $\theta_{ij}$ and $s_{ij}$ stated above.

The corresponding Drinfel'd twist is 
\beq\label{geo21}
F_\sigma=\sum_{\vec{m},\vec{m}'}\sigma(\vec{m},\vec{m}')\mathbb{P}_{\vec{m}}\otimes\mathbb{P}_{\vec{m}'}
\eeq

Note that
\beq
|\sigma(\vec{m},\vec{m}')|=1,\quad F^{-1}_\sigma=F_{\bar{\sigma}}\quad.
\eeq

If $\epsilon$ is the counit, then there is the normalisation condition \cite{majid}
\beq
(\epsilon\otimes\I)F_\sigma=(\I\otimes\epsilon)F_\sigma=\I\quad.
\eeq
Here $\epsilon$ is the map to the ``trivial'' representation, so $\vec{m}$ and $\vec{m}'$ become $\vec{0}$ (mod $\vec{n}=(n_1,..,n_k)$) under $\epsilon$ and $\epsilon(\mathbb{P}_{\vec{m}})=\delta_{\vec{m},\vec{0}}$, $\epsilon(\mathbb{P}_{\vec{m}'})=\delta_{\vec{m}',\vec{0}}$. Since $\sum_{\vec{m}}\mathbb{P}_{\vec{m}}=\sum_{\vec{m}'}\mathbb{P}_{\vec{m}'}=\I$, the above requirement is fulfilled by (\ref{geo21}).

Next we show that the symmetric factor with $s_{ij}$ can be eliminated by requiring that the twist preserves the adjoint operation. 

For the twist $F_\sigma$ above, the dressed annihilation and creation operators are
\bea
a_{\vec{m}}=\sum_{\vec{m}'}c_{\vec{m}}\e^{-\frac{i}{2}m_i(\theta_{ij}+s_{ij})m'_j}\mathbb{P}_{\vec{m}'}\\
a^*_{\vec{m}}=\sum_{\vec{m}'}c^\dag_{\vec{m}}\e^{\frac{i}{2}m_i(\theta_{ij}+s_{ij})m'_j}\mathbb{P}_{\vec{m}'}
\eea
where $*$ denotes that it is not necessarily the adjoint $^\dag$ of $a_{\vec{m}}$, and we have used the fact that $a^*_{\vec{m}}$ transforms by the representation $-\vec{m}$.

Now
\beq
a^\dag_{\vec{m}}=\Big(\sum_{\vec{m}'}\e^{\frac{i}{2}m_i(\theta_{ij}+s_{ij})m'_j}\mathbb{P}_{\vec{m}'}\Big)c^\dag_{\vec{m}}
\eeq
The prefactor is the unitary operator $U_{\vec{m}}$ representing the element
\beq
\times_j\ \e^{\frac{i}{2}m_i(\theta_{ij}+s_{ij})}
\eeq
in $A$. Hence
\beq
U_{\vec{m}}c^\dag_{\vec{m}}U_{\vec{m}}^{-1}=\e^{\frac{i}{2}m_i(\theta_{ij}+s_{ij})m_j}c^\dag_{\vec{m}}=\e^{\frac{i}{2}m_is_{ij}m_j}c^\dag_{\vec{m}}
\eeq
since $\theta_{ij}=-\theta_{ji}$. Thus
\beq
a^\dag_{\vec{m}}=\e^{\frac{i}{2}m_is_{ij}m_j}a^*_{\vec{m}}\quad.
\eeq

The requirement 
\beq
a^*_{\vec{m}}=a^\dag_{\vec{m}} \, .
\eeq
imposes the constraint 
\beq\label{constraint}
\e^{\frac{i}{2}m_is_{ij}m_j}=1
\eeq
From this we can infer that $s_{ij} = 0$ mod $4\pi/m_{ij}$ where $n_i/m_{ij}$, $n_j/m_{ij}\in \Z$.
For example the successive choices $\vec{m} = (1,{\bf 0}), (0,1,\vec{0}), (1,1, \vec{0})$, shows that $s_{ij} = 0$ if
$i,j \leq 2$.  Thus (\ref{constraint}) reduces $\sigma$ to
\beq
\sigma(\vec{m},\vec{m}')=\e^{-\frac{i}{2}m_i\theta_{ij}m'_j}
\eeq

It thus appears that our previous considerations are general for associative spacetime algebras.

\subsection{Nonabelian Generalisations of Drinfel'd Twists}

We now discuss nonabelian generalisations of the above considerations. They generally lead to quasi-Hopf algebras based on $D^\infty/D^\infty_0$ as the symmetry algebras and non-associative spacetimes. 

Here is an approach to such a generalisation. Let us consider the following nested groups:
\beq\label{geo22}
D^\infty/D^\infty_0\equiv G_0\supset G_1\supset...\supset G_N=A
\eeq
Here $A=\times^k_{i=1}\Z_{n_i}$ is the maximal abelian subgroup of $G_0$ (quotiented by factors of $\Z$)  while the rest, $G_k$ for $k<N$, can be nonabelian. The chain is supposed to be such that there exists an orthonormal basis $\{b^{(\vec{\varrho})}\}$ $\Big(\vec{\varrho}=(\varrho_0,\varrho_1,...,\varrho_N)\Big)$ for the vector space $V^{(\varrho_0)}$ for the IRR $\varrho_0$ of $G_0$ where $b^{(\vec{\varrho})}$ is a vector in the representation space for the IRR $\varrho_j$ of $G_j$. In this notation, $\varrho_N$= our previous $\vec{m}$. Thus the chain (\ref{geo22}) leads to a complete system of labels for the basis vectors.

Let $\mathbb{P}_{\vec{\varrho}}$ be the projector to the space $\mathbb{C}b^{(\vec{\varrho})}$:
\beq
\mathbb{P}_{\vec{\varrho}}\  b^{(\vec{\varrho})}=b^{(\vec{\varrho})}\quad.
\eeq
Then
\beq
\mathbb{P}_{\vec{\varrho}}\mathbb{P}_{\vec{\varrho}'}=\delta_{\vec{\varrho},\vec{\varrho}'}\mathbb{P}_{\vec{\varrho}},\quad\sum_{\vec{\varrho}}\mathbb{P}_{\vec{\varrho}}=\I\quad.
\eeq

Let $\vec{\varrho}_\epsilon$ label the IRR associated with the counit $\epsilon$. Then
\beq
\epsilon(\mathbb{P_{\vec{\varrho}}})=\delta_{\vec{\varrho},\vec{\varrho}_\epsilon}
\eeq

Now consider
\beq\label{geo23}
F_\sigma=\sum_{\vec{\varrho},\vec{\varrho}'}\sigma(\vec{\varrho},\vec{\varrho}')\mathbb{P}_{\vec{\varrho}}\otimes\mathbb{P}_{\vec{\varrho}'},\quad\sigma(\vec{\varrho},\vec{\varrho}')\in\mathbb{C}\quad.
\eeq
We plan to use $F_\sigma$ as the Drinfel'd twist. Its realization used to deform the $\star$-product of functions will be indicated as usual as $\mathscr{F}_\sigma$. It involves the realization of $\mathbb{P}_{\vec{\varrho}}$'s on functions which again we will call $\PR_{\vec{\varrho}}$'s. The Drinfel'd twist of the coproduct as in (\ref{geo7}), requires $F_\sigma$ to be invertible so that
\beq
\sigma(\vec{\varrho},\vec{\varrho}')\neq0\quad{\rm for\ any\ \vec{\varrho},\vec{\varrho}'}\quad.
\eeq
For then,
\beq
F^{-1}_\sigma=\sum_{\vec{\varrho},\vec{\varrho}'}\frac{1}{\sigma(\vec{\varrho},\vec{\varrho}')}\mathbb{P}_{\vec{\varrho}}\otimes\mathbb{P}_{\vec{\varrho}'}\quad.
\eeq

The next requirement on $F_\sigma$ is the normalisation condition
\beq
(\epsilon\otimes\I)F_\sigma=(\I\otimes\epsilon)F_\sigma=\I\quad.
\eeq
In view of (\ref{geo23}), this requires that
\beq\label{geo24}
\sigma(\vec{\varrho}_\epsilon,\vec{\varrho})=\sigma(\vec{\varrho},\vec{\varrho}_\epsilon)=1\quad.
\eeq

According to Majid \cite{majid}, there is no further requirement on $F_\sigma$ if quasi-Hopf algebras are acceptable. The spacetime algebra with its star product governed by $F_\sigma$ as in previous sections is then its module algebra which is generally nonassociative (with an associator) \cite{MackSh1,MackSh2,MackSh3}. It is associative only if its symmetry algebra is Hopf.

The spacetime orthonormal basis is now denoted by $b^{(\pm)}_{\vec{\varrho}}$ instead of by $f^{(\pm)}_{\vec{m}}$ while the twisted quantum field is written as
\bea\label{geo26}
&\varphi_\theta=\sum_{\vec{\varrho},\vec{\varrho}'}\Big[a_{\vec{\varrho}}\ b^{(+)}_{\vec{\varrho}}+a^*_{\vec{\varrho}}\ b^{(-)}_{\vec{\varrho}}\Big]&\\
&a_{\vec{\varrho}}=\sum_{\vec{\varrho}'}c_{\vec{\varrho}}\ \sigma(\vec{\varrho},\vec{\varrho}')\mathbb{P}_{\vec{\varrho}'}&\\
&a^*_{\vec{\varrho}}=\sum_{\vec{\varrho}'}c^\dag_{\vec{\varrho}}\ \bar{\sigma}(\vec{\varrho},\vec{\varrho}')\mathbb{P}_{\vec{\varrho}'}&
\eea
where $c_{\vec{\varrho}}$, $c^\dag_{\vec{\varrho}}$ are the untwisted annihilation and creation operators. 

Unitarity requires that
\beq
a^*_{\vec{\varrho}}=a^\dag_{\vec{\varrho}}=\sum_{\vec{\varrho}'}\bar{\sigma}(\vec{\varrho},\vec{\varrho}')\mathbb{P}_{\vec{\varrho}'}c^\dag_{\vec{\varrho}}=\sum_{\vec{\varrho}'}c^\dag_{\vec{\varrho}}\ \bar{\sigma}(\vec{\varrho},\vec{\varrho})\ \bar{\sigma}(\vec{\varrho},\vec{\varrho}')\mathbb{P}_{\vec{\varrho}'}\quad.
\eeq
Hence we have also
\beq\label{geo25}
\sigma(\vec{\varrho},\vec{\varrho})=1\quad.
\eeq

Thus it appears that we have an approach to a quantum field theory if the normalisation condition (\ref{geo24}) and the unitary condition (\ref{geo25}) are fulfilled. 

If $\PR_{\vec{\varrho}}$ is the projector on the space of functions to the IRR $\vec{\varrho}$, the twisted field can be written without a mode expansion:
\beq\label{geo30}
\varphi_\theta=\sum_{\vec{\varrho},\vec{\varrho}'}\sigma(\vec{\varrho},\vec{\varrho}')\Big(\PR_{\vec{\varrho}}\varphi_0\Big)\mathbb{P}_{\vec{\varrho}'}\quad.
\eeq
It is then easily verified that the dressed field (\ref{geo30}) coincides with (\ref{geo26}) and it has the self-reproducing property:
\beq
\varphi_\theta\star\varphi_\theta=\sum_{\vec{\varrho},\vec{\varrho}'}\sigma(\vec{\varrho},\vec{\varrho}')\Big(\PR_{\vec{\varrho}}\varphi_0^2\Big)\mathbb{P}_{\vec{\varrho}}\quad.
\eeq
But there is in general no associativity:
\beq
(\varphi_\theta\star\varphi_\theta)\star\varphi_\theta\neq\varphi_\theta\star(\varphi_\theta\star\varphi_\theta)\quad.
\eeq

Such quantum fields merit study. They seem to lead to Pauli principle violations with testable experimental consequences. We will elaborate on this remark elsewhere.

\section{On Rings and Their Statistics (Motion) Groups}

A theoretical approach to the investigation of statistics of a system of identical constituents is based on the properties of the fundamental group of its configuration space. For $N$ spinless identical particles in a Euclidean space of three or more dimensions, for example, this group is known to be the permutation group $S_N$ \cite{book}. There is furthermore an orderly method for the construction of a distinct quantum theory for each of its unitary irreducible representations (UIRs). As these theories describe bosons, fermions and paraparticles according to the choice of the representation, the study of the fundamental group leads to a comprehensive account of the possible statistics of structureless particles in three or more dimensions.

It has been appreciated for some time that the statistical possibilities of a particle species confined to the plane $\mathbb{R}^2$ can be quite different from those in three or more dimensions. This is because the fundamental group for $N$ identical spinless particles in a plane is not $S_N$. It is instead an infinite group $B_N$, known as the braid group. Since $S_N$ is a factor group of $B_N$, and hence representations of $S_N$ are also those of $B_N$, it is of course possible to associate Bose, Fermi or parastatistics with a particle species in a plane. But since $B_N$ has many more UIRs which are not UIRs of $S_N$, there are also several possibilities for exotic planar statistics. One such possibility of particular interest, for instance, is that of fractional statistics, which is of importance in the context of fractional quantum Hall effect.

As we discussed earlier, it was pointed out some time ago that configuration space with unusual fundamental groups, and hence exotic statistical possibilities, occur not merely for point particles on a plane, but also for topological geons. It was also emphasized elsewhere \cite{Anez} that there are many remarkable properties associated with the quantum version of geons, such as the failure of the spin-statistics connection and the occurrence of states in three spatial dimensions which are not bosons, fermions or paraparticles.

In \cite{Kauff,Islam} the investigation of exotic statistics was continued by examining another system of extended objects, namely a system of identical closed strings assumed to be unknots and imbedded in three spatial dimensions. Using known mathematical results on motion groups \cite{motion}, it was shown that the fundamental group of the configuration space of two or more such strings is not the permutation group either. It is instead an infinite non-Abelian group which bears a certain resemblance to the gravitational fundamental groups  mentioned a moment ago. It was further shown that quantum strings as well may not be characterized by permutation group representations. Thus they may not obey Bose, Fermi or parastatistics. They may also fail to obey the familiar spin-statistics connection. 

Thus identical geons and identical knots share certain topological properties. For this reason, in this section we briefly examine the statistics of identical unknots. We here consider only the configuration spaces of one and two unknots and their fundamental groups.

We denote the configuration space of $N$ unknots in $\R^3$ as $\mathcal{Q}^{(N)}$ and consider $N=1$ and 2. These unknots can be unoriented or oriented. These cases will be discussed separately.

\subsection{{\it The case of one unoriented unknot}}

An unoriented unknot is a map of a circle $S^1$ into $\R^3$ where the image is the unoriented unknot. That means the following:

\begin{itemize}

\item[a)] It can be deformed to the standard map where the image is say the circle $\{(x,y,0):\sum x^2+y^2=1\}$ in the 1-2 plane. (Here we chose the flat metric $\delta_{ij}$).

\item[b)] Two maps which differ by an orientation reversal of $S^1$ are identified.

\end{itemize}

Intuitively, an unknot is a closed loop deformable to the above standard loop. 

The configuration space $\mathcal{Q}^{(1)}$ of the unknot consists of all such maps. 

We now consider the fundamental group $\pi_1(\mathcal{Q}^{(1)})$. 

The construction of $\pi_1(\mathcal{Q}^{(1)})$ involves the choice of a fixed (``base'') point $\bar{q}$ in $\mathcal{Q}^{(1)}$. As $\mathcal{Q}^{(1)}$ is the space of maps from $S^1$ to $\R^3$, $\bar{q}$ in this case is one particular choice of such maps. If $\mathcal{Q}^{(1)}$ is connected, as is the case for us, it can be any point $\bar{q}$ of $\mathcal{Q}^{(1)}$. The resultant group $\pi_1(\mathcal{Q}^{(1)},\bar{q})$, where we have put in the base point $\bar{q}$ in the notation for the fundamental group, does not depend on $\bar{q}$. So we can talk of $\pi_1(\mathcal{Q}^{(1)})$ and omit $\bar{q}$. 

But there is no canonical isomorphism between $\pi_1(\mathcal{Q}^{(1)};\bar{q})$ and $\pi_1(\mathcal{Q}^{(1)};\bar{q}')$ with 
$\bar{q}\neq\bar{q}'$. Any isomorphism depends on the choice of the path from $\bar{q}$ to $\bar{q}'$ \cite{book}.

\begin{figure} 
\includegraphics[scale=.5]{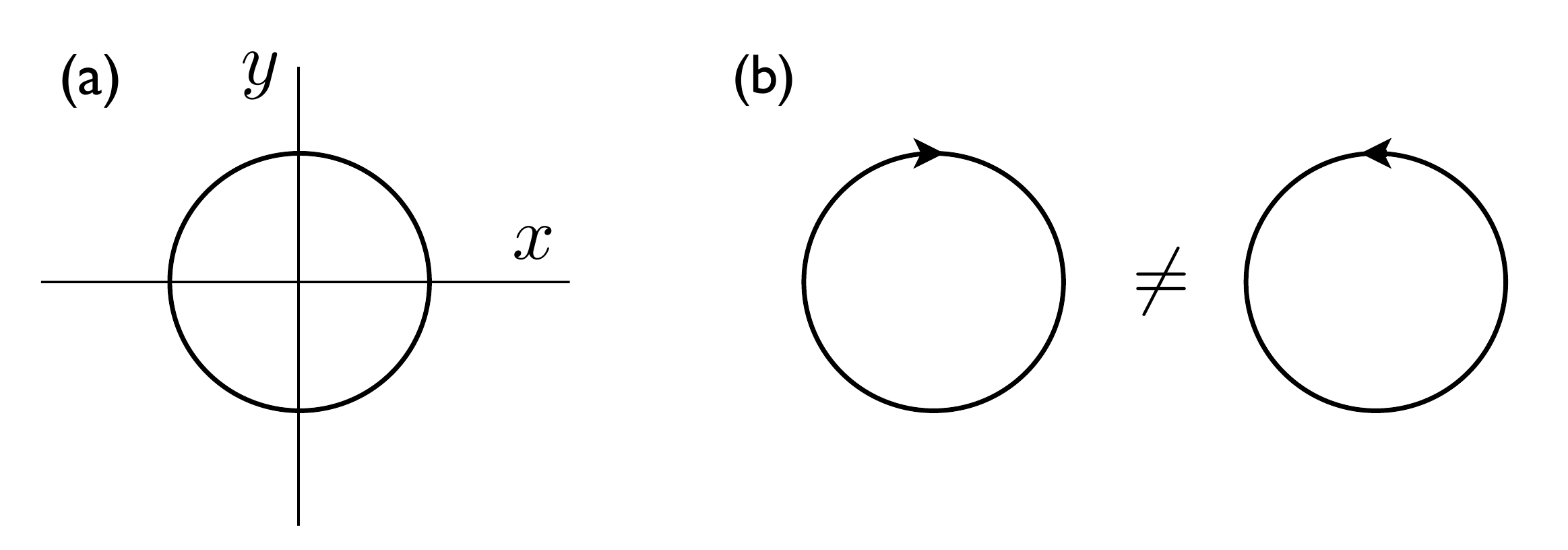} 
\caption{(a) Circle of unit radius centered at the origin. (b) Two oriented inequivalent knots.}
\label{fig5}
\end{figure}

For $\bar{q}$, we can for convenience choose the flat metric $\delta_{ij}$ in $\R^3$ as we did above, and choose $\bar{q}$ to be a circle of unit radius centered at the origin in the $x-y$ plane as in Fig. \ref{fig5} (a).

Consider rotating this figure by $\pi$ around the $x$-axis. It maps $\bar{q}$ to $\bar{q}$ and creates a loop T in $\mathcal{Q}^{(1)}$ as the rotation evolves from 0 to $\pi$. The loop cannot be deformed to a point, the point loop based at $\bar{q}$. So [T], the homotopy class of this loop is a non-trivial element of $\pi_1(\mathcal{Q}^{(1)})$. 

Rotating $\bar{q}$ around another axis $\hat{n}$ ($\hat{n}\cdot\hat{n}=1$) through the origin generates a loop which however is homotopic to T: just consider the sequence of loops got by rotating $\hat{n}$ to the $x$-axis $\hat{i}$ to this result.

By repeating T $k$-times, we get a $k\pi$ rotation loop call it T$^k$, of $\bar{q}$. If $J_1$ is the angular momentum of $SO(3)$, then
\beq
\{\e^{i\theta J_1}\ :\ 0\leq\theta\leq2\pi\}
\eeq
is a 2$\pi$-rotation loop in $SO(3)$ and T$^2$ is just $\{\e^{i\theta J_1}\bar{q}\ :\ 0\leq\theta\leq2\pi\}$.

But this loop can be deformed to a point. For consider the sequence of loops
\beq
\{\e^{i\theta\hat{n}\vec{J}}\bar{q}:\ 0\leq\theta\leq2\pi,\ \hat{n}\cdot\hat{n}=1\}
\eeq
as $\hat{n}$ varies from (1,0,0) to (0,0,1). The starting loop is T, the final loop is a point. Thus [${\rm T}^2$]=$e$.

We thus see that
\beq
\pi_1(\mathcal{Q}^{(1)})=\mathbb{Z}_2=\langle[T],[{\rm T}^2]=e\rangle
\eeq

\subsection{{\it The case of the oriented unknot}}

In this case we drop the identification $b)$ above so that there is an arrow attached to the unknot like in Fig. \ref{fig5} (b).
Otherwise, its configuration space $\mathcal{Q}^{(1)}$ is defined as above. 

As for $\pi_1(\mathcal{Q}^{(1)})$, the base point $\bar{q}$ is as above, but there is now an arrow on the circle in Fig. \ref{fig5} (b). Hence the curve
\beq
T=\langle\e^{i\theta J_1}\bar{q}\ :\ 0\leq\theta\leq\pi\rangle
\eeq
does not close (is not a loop). We conclude that
\beq
\pi_1(\mathcal{Q}^{(1)})=\{e\}
\eeq

\begin{figure} 
\includegraphics[scale=.4]{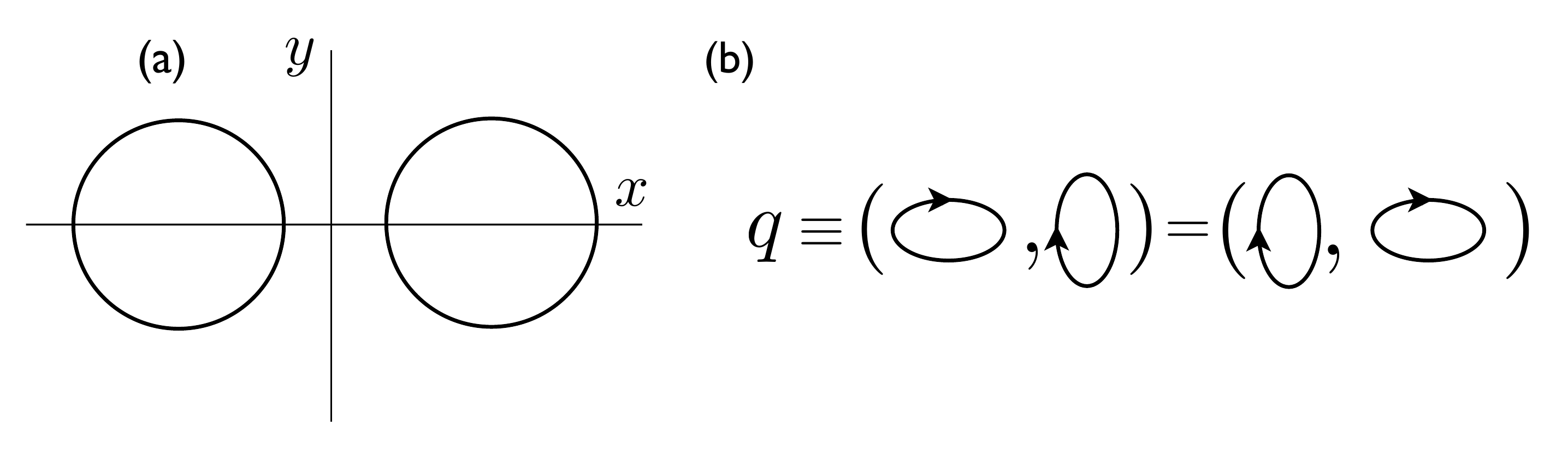} 
\caption{(a) The configuration space of two identical unoriented unknots.  (b) Pair of unordered identical knots. }
\label{fig7}
\end{figure}

\subsection{{\it The case of two identical unoriented unknots}}

Its configuration space $\mathcal{Q}^{(2)}$ can be informally described (see Fig.  \ref{fig7}) (a) as follows: A point $q\in\mathcal{Q}^{(2)}$ consists now of 2 unlinked unknots in $\mathbb{R}^3$.
The pair is unordered as the knots are ``identical'', see Fig. \ref{fig7} (b).  
This requirement is as for identical particles \cite{book}.  

For $\bar{q}$, using our flat metric, we choose two circles of unit radius on the $x-y$ plane centered in $\pm2$.

The discussion of identical unknots here is to be compared with the corresponding discussion of identical geons \cite{Anez}.

We can now recognize the following elements of $\pi_1(\mathcal{Q}^{(2)})$:

\begin{figure} 
\includegraphics[scale=.4]{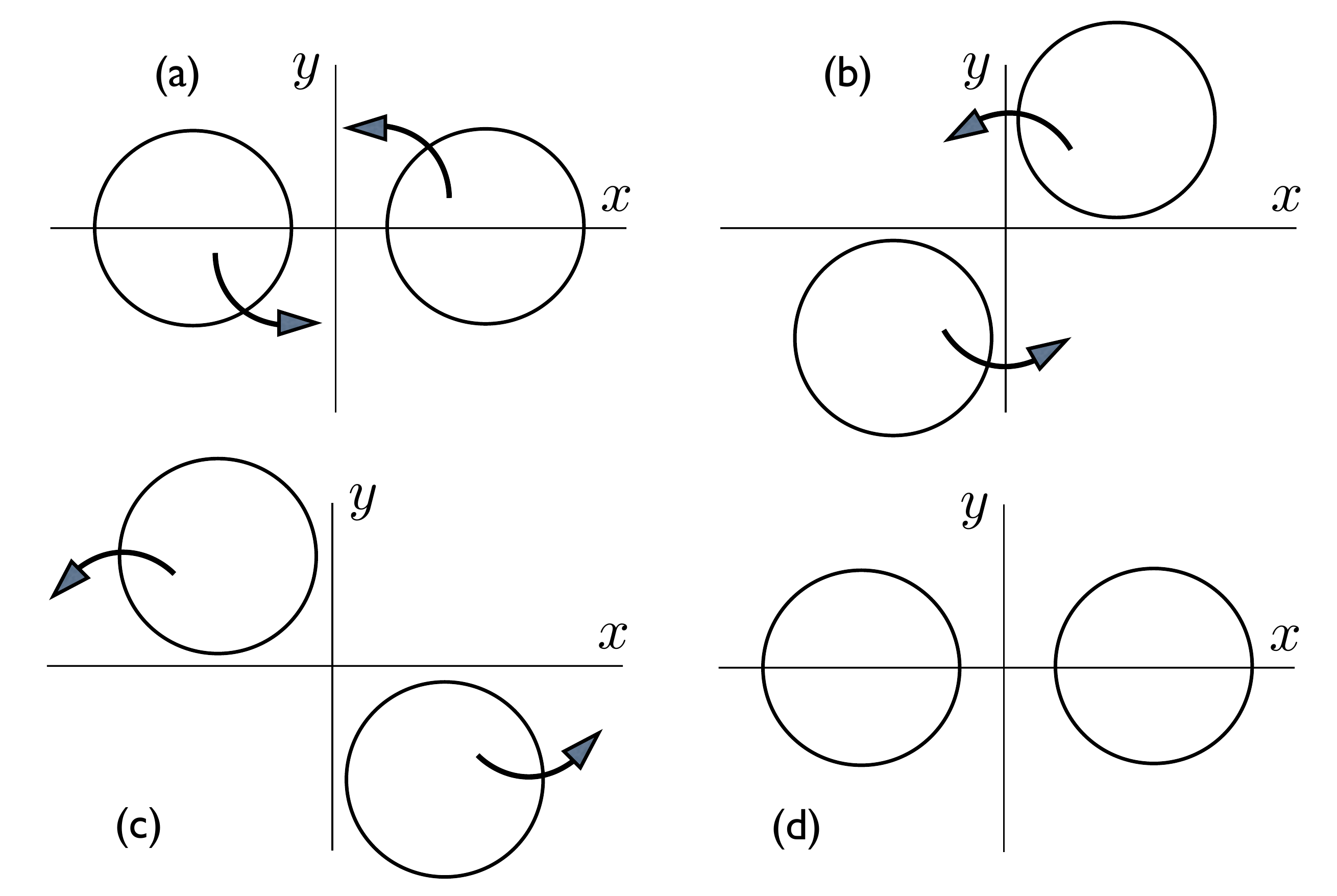} 
\caption{The loop E defining ``exchange''.}
\label{fig9}
\end{figure}

\begin{itemize}

\item {\em Exchange} [E]:  The loop E defining ``exchange'' rotates $\bar{q}$ from 0 to $\pi$ around third axis:
\beq
{\rm E}=\langle\e^{i\theta J_3}\bar{q}\ :\ 0\leq\theta\leq\pi\rangle\quad.
\eeq
Its evolving pictures are depicted in Fig. \ref{fig9}.

There are standard proofs that 
\beq\label{eq}
{\rm E}^2\equiv\langle\e^{i\theta J_3}\bar{q}\ :\ 0\leq\theta\leq2\pi\rangle
\eeq
is deformable to $e$ and that the loop with $\theta\to-\theta$ in (\ref{eq}) is homotopic to E. 

The homotopy class [E] of E in $\pi_1(\mathcal{Q}^{(2)})$ is the exchange. The corresponding group is $S_2$.

\item {\it The $\pi$-rotations} $[{\rm T}^{(1)} ], [ {\rm T}^{(2)}]$.

The loop T$^{(1)}$ rotates the ring 1 (on left) by $\pi$ around $x$-axis,  T$^{(2)}$ does so for the ring 2 on right. They are inherited from $\mathcal{Q}^{(1)}$ and generate the elements [T$^{(i)}$] in $\pi_1(\mathcal{Q}^{(2)})$. Clearly
\beq
[{\rm E}][{\rm T}^{(1)}][{\rm E}^{-1}]=[{\rm T}^{(2)}],\quad[{\rm E}][{\rm T}^{(2)}][{\rm E}^{-1}]=[{\rm T}^{(1)}]
\eeq
where the products in $\pi_1(\mathcal{Q}^{(2)})$ are as usual defined by concatenation of loops in $\mathcal{Q}^{(2)}$.

[T$^{(1)}$] and [T$^{(2)}$] commute.

\begin{figure} 
\includegraphics[scale=.4]{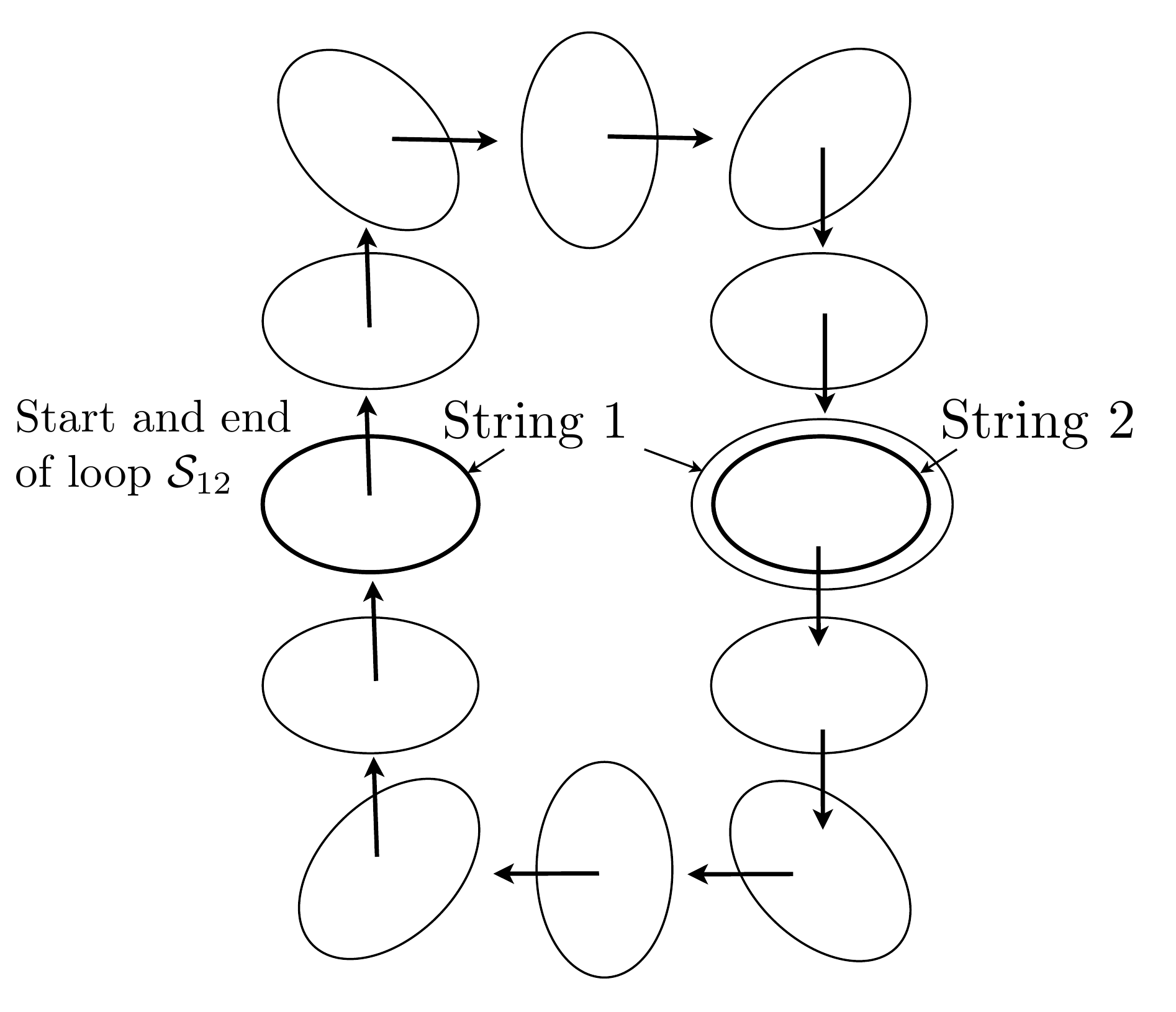} 
\caption{The ``slide'' loop $\mathcal{S}_{12}$.}
\label{fig10}
\end{figure}

\item {\it The Slide:}  Let us first consider the loop $\mathscr{S}_{12}$ or the slide $[\mathscr{S}_{12}]$ of 2 through 1. Figure \ref{fig10} above explains the loop:

The homotopic class $[\mathscr{S}_{12}]$ of $\mathscr{S}_{12}$ is the slide of 2 through 1. 

The slide $[\mathscr{S}_{21}]$ of 1 through 2 is similarly defined. We can show that
\bea\label{eq2}
[{\rm E}][\mathscr{S}_{12}][{\rm E}]^{-1} = [\mathscr{S}_{21}]\quad,\\\label{eq3}
[{\rm E}][\mathscr{S}_{21}][{\rm E}]^{-1} = [\mathscr{S}_{12}]\quad.
\eea
The full group $\pi_1(\mathcal{Q}^{(2)})$ is thus generated by [T$^{(1)}$], [T$^{(2)}$], [E], [$\mathscr{S}_{12}$], [$\mathscr{S}_{21}$]  with the relations 
\beq
[{\rm T}^{(1)}]^2 = [{\rm E}] ^2 = [{\rm T}^{(2)}]^2 = e ,  [{\rm E}][{\rm T}^{(1)}][{\rm E}]^{-1} = [{\rm T}^2],   [{\rm T}^{(1)}][{\rm T}^{(2)}] = [{\rm T}^{(2)}][{\rm T}^{(1)}]  \,,
\eeq
 and (\ref{eq2},\ref{eq3}). There are no further relations.

If $\mathscr{S}$ is the group that [$\mathscr{S}_{ij}$] generate, we have the semi-direct product structure
\beq\label{eq4}
\pi_1(\mathcal{Q}^{(2)})=\left\{\mathscr{S}\rtimes\Big(\pi_1(\mathcal{Q}^{(1)})\times\pi_1(\mathcal{Q}^{(1)})\Big)\right\}\rtimes S_2
\eeq

Here $G_1\rtimes G_2$ is the semi-direct product of $G_1$ and $G_2$ with $G_1$ being the invariant subgroup. Also $\pi_1(\mathcal{Q}^{(1)})$ acts trivially on $\mathscr{S}$.

Eq. (\ref{eq4}) is to be compared with the corresponding equation for the mapping class group $D^{(2)\infty}/D^{(2)\infty}_0$ of two identical geons \cite{MCG1,MCG2} if $\mathscr{S}$ is its group of slides,
\beq
D^{(2)\infty}/D^{(2)\infty}_0=\left\{\mathscr{S}\rtimes\Big(D^{(1)\infty}/D^{(1)\infty}_0\times D^{(1)\infty}/D^{(1)\infty}_0\Big)\right\}\rtimes S_2
\eeq

\end{itemize}

\subsection{\it The case of two identical oriented unknots}

Orienting the knots reduces $\pi_1(\mathcal{Q}^{(1)})$ to $\{e\}$. With that in mind, we can repeat the above discussion (with $\bar{q}$ chosen analogously to above) to find
\beq
\pi_1(Q^{(2)})=\mathscr{S}\rtimes S_2\quad.
\eeq
Some discussion about the quantum theory of these unknots and their unusual statistical features can be found in \cite{Kauff,Islam}.

\section{Final Remarks}

Topological geons were discovered by Friedman and Sorkin. In this paper, we have developed an approach for twisting the spacetimes of topological geons. The twists we consider are localised on the geons and are of the order of Planck scales. They lead to spacetime noncommutativity only on these scales as required by Doplicher, Fredenhagen and Roberts \cite{Doplicher}. When geons emerge from asymptotically flat spacetimes, these twists do not seem to affect the Poincar\'e symmetry for the Wightman functions \cite{mangano}. But that need not be the case for scattering amplitudes \cite{Scatt}.  

A significant new contribution of this paper is the extension of Drinfel'd twists to discrete groups and the construction of the associated quantum field theories.

Generic Drinfel'd twists lead to nonassociative spacetimes (such as in \cite{Nonass}) and their quantum fields. For geons, in our approach, such nonassociativity is manifest only at Planck scales. We have reasons to think that Pauli principle violation is a feature of these quantum field theories. While we do not elaborate on this point here, we do plan to study this issue later.

\acknowledgements
It is a pleasure for Balachandran, Marmo and Martone to thank Alberto Ibort and the Universidad Carlos III de Madrid for their wonderful hospitality and support.

The work of Balachandran and Martone was supported in part by DOE under the grant number DE-FG02-85ER40231 by the Department of Science and Technology (India) and by the Institute of Mathematical Sciences, Chennai. We thank Professor T. R. Govindarajan for his very friendly hospitality at the Institute of Mathematical Sciences, Chennai. Balachandran was also supported by the Department of Science and Technology, India.

A. Ibort would like to acknowledge the partial support from MICINN research project MTM2007-62478 and QUITEMAD.

\bibliographystyle{apsrmp}

\end{document}